# Planetary-scale variations in winds and UV brightness at the Venusian cloud top: Periodicity and temporal evolution

**Masataka Imai[1], Toru Kouyama[1], Yukihiro Takahashi[2], Atsushi Yamazaki[3], Shigeto Watanabe[4], Manabu Yamada[5], Takeshi Imamura[6], Takehiko Satoh[3], Masato Nakamura[3], Shin-ya Murakami[3], Kazunori Ogohara[7], Takeshi Horinouchi[8]**

[1]Artificial Intelligence Research Center, Advanced Industrial Science and Technology, 2-4-7, Aomi, Koto-ku, Tokyo 135-0064, Japan.

[2]Planetary and Space Group, Department of Cosmosciences, Graduate School of Science, Hokkaido University, Kita 10-Nishi 8, Kita-ku, Sapporo, Hokkaido 060-0810, Japan.

[3]Institute of Space and Astronautical Science, Japan Aerospace Exploration Agency, 3-1-1, Yoshinodai, Chuo-ku, Sagamihara, Kanagawa 252-5210, Japan.

[4]Space Information Center, Hokkaido Information University, 59-2, Nishinopporo, Ebetsu, Hokkaido 069-8585, Japan.

[5]Planetary Exploration Research Center, Chiba Institute of Technology, 2-17-1, Tsudanuma, Narashino, Chiba 275-0016, Japan.

[6]Graduate School of Frontier Sciences, The University of Tokyo, 5-1-5, Kashiwanoha, Kashiwa, Chiba 277-8561, Japan.

[7]School of Engineering, The University of Shiga Prefecture, 2500, Hassaka-cho, Hikone, Shiga 522-8533, Japan.

[8]Faculty of Environmental Earth Science, Hokkaido University, Kita 10-Nishi 5, Kita-ku, Sapporo, Hokkaido 060-0810, Japan.

Corresponding author: Masataka Imai (mstk-a.imai@aist.go.jp)

**Key Points:**

- Periodic variations in winds and UV brightness observed by Akatsuki were analyzed to study planetary-scale waves at the Venusian cloud top

- The temporal evolution of Rossby and Kelvin waves was captured for the first time

- The relation between the waves in the winds and UV brightness is described

## Abstract

Planetary-scale waves at the Venusian cloud-top cause periodic variations in both winds and ultraviolet (UV) brightness. While the wave candidates are the 4-day Kelvin wave and 5-day Rossby wave with zonal wavenumber 1, their temporal evolutions are poorly understood. Here we conducted a time series analysis of the 365-nm brightness and cloud-tracking wind variations, obtained by the UV Imager onboard the Japanese Venus Climate Orbiter Akatsuki from June to October 2017, revealing a dramatic evolution of planetary-scale waves and corresponding changes in planetary-scale UV features. We identified a prominent 5-day periodicity in both the winds and brightness variations, whose phase velocities were slower than the dayside mean zonal winds (or the super-rotation) by >35 m $s_{-1}$. The reconstructed planetary-scale vortices were nearly equatorially symmetric and centered at ~35° latitude in both hemispheres, which indicated that they were part of a Rossby wave. The amplitude of winds variation associated with the observed Rossby wave packet were amplified gradually over ~20 days and attenuated over ~50 days. Following the formation of the Rossby wave vortices, brightness variation emerges to form rippling white cloud belts in the 45°–60° latitudes of both hemispheres. ~3.8-day periodic signals were observed in the zonal wind and brightness variations in the equatorial region before the Rossby wave amplification. Although the amplitude and significance of the 3.8-day mode were relatively low in the observation season, this feature is consistent with a Kelvin wave, which may be the cause of the dark clusters in the equatorial region.

## Plain Language Summary

The Earth's twin planet Venus is mysterious for the fast atmosphere circulation called as the super-rotation. The cloud top atmosphere rotates around the planet with 100 m $s_{-1}$, which corresponds to ~4-day circulation. There are many types of atmospheric waves, which are crucial for understanding how the wind blows on the planet. In this study, we analyzed one thousand Venus ultraviolet (UV) images taken by Japanese Venus Climate Orbiter Akatsuki and firstly captured the continuous temporal evolution of a planetary-scale wave with a period of five days. The 5-day wave had large equatorially symmetric vortices at the cloud top. According to the wave evolution, the shape of the dark UV cloud (absorption) features dramatically changed over the planet. Since the UV absorption is important for the radiative energy balance, we open the door for the future investigations of the long-term impacts of planetary-scale waves to the atmospheric dynamics and chemistry.

## 1 Introduction

The Venusian cloud-top level, which locates ~65–70 km altitude, is dominated by fast planetary-scale westerly winds. The zonal wind reaches ~100 m s–1, and this prominent phenomenon is named the super-rotation. The successful Venus Express (2006–2014) exploration missions of the European Space Agency acquired extensive data on the Venusian atmosphere, and the Venus Monitoring Camera (VMC) captured a large number of high-contrast ultraviolet (UV) cloud images at the 365-nm wavelength for unknown absorbers (Markiewicz et al., 2007; Titov et al., 2008, 2012). Using a cloud-tracking technique (e.g., recently used by Khatuntsev et al., 2013; Kouyama et al., 2013; Ikegawa and Horinouchi, 2016), the large variations in averaged zonal wind of ~100 ± 20 m s–1 in the equatorial region were found during the VMC observation seasons, with an estimated time scale of about one Venusian year (224 Earth days) or longer (e.g., Kouyama et al. (2013) suggested ~255 days).

The prominent, planetary-scale UV features hold an important clue to understand the atmospheric dynamics of Venus. These features are constructed from the longitudinal bright and dark absorption features in the low-latitude regions and the dark stream bands in the mid-latitude regions. They are sometimes called "Y-features" (or "ψ-features") because they look like a sideways Y (or ψ). Belton et al. (1976) noted that the migration speed of these prominent planetary-scale UV features is inconsistent with the mean cloud-tracked background zonal winds. Del Genio and Rossow (1990) investigated the relationship between the phase velocity of the propagating wave, which was measured from the cyclic period of the UV brightness variation, and mean zonal winds, observed by the Orbiter Cloud Photo-Polarimeter aboard Pioneer Venus in the 1980s. This provided the first observational results demonstrating variations in planetary-scale waves, and also indicating two types of waves. The first could be a Kelvin wave, with a ~4-day (day = Earth day) period, that propagated westward ~15 m s–1 faster than the zonal mean wind in the equatorial region. The second was considered a Rossby wave, with a ~5-day period, which had a retrograde propagating mode against the zonal mean winds and a phase velocity of >30 m s–1 in the mid-latitude region. Covey and Schubert (1982) showed that Kelvin-like waves and Rossby-like waves are excited as preferred modes for cloud-level forcing by studying the linear response of a model Venus atmosphere to external forcing over broad frequencies. Smith et al. (1993) argued that the Kelvin wave could be a global mode of the Venus atmosphere and it could be generated by the cloud heating below the cloud top level. The roles of these waves in the formation of the planetary-scale Y feature was investigated by Peralta et al. (2015) and Nara et al. (2019).

Kouyama et al. (2013, 2015) conducted further investigations of these planetary-scale waves using the Venus Express/VMC cloud-tracked wind data. Zonal and meridional wind variations, with ~4–5-day periods, were analyzed in 13 observation epochs between May 2006 and December 2011, with Rossby-like waves often observed as a dominant mode. Furthermore, Kelvin-like waves clearly appeared from July to September 2007, when the background zonal wind was at its minimum (~90 m s–1) in the VMC observations. Their numerical study showed that variations in the background zonal wind may determine which type of planetary-scale wave can propagate in the vertical direction. Since dissipation of either Kelvin or Rossby waves at the cloud-top level should be the reason for accelerating or decelerating zonal wind, their results highlighted the importance of the appearance and disappearance of planetary-scale waves on Venusian atmospheric dynamics.

Previous Venus exploration missions, such as Pioneer Venus and Venus Express, provided long-term UV monitoring data from the Venusian cloud top. However, since the orbital planes of the spacecraft were approximately fixed in inertial space, the ~4–5-day periodic variations in UV brightness and winds within sub-Venusian-year time scales were obscured by limitations of the continuous dayside observations. Therefore, the temporal changes in sub-Venusian-year periodic variations were not studied in detail. Our previous work (Imai et al., 2016) implemented long-term monitoring of the rotation period of planetary-scale UV features using a ground-based telescope. Two significant periodicities, at 5.1- and 3.5-day, should be manifestations of these planetary-scale waves, which were subjected to temporal variations within several months. However, the temporal evolution of these variations still remains largely unknown, due to the accuracy of ground-based measurements, since it is likely too low for investigating variability in the periodicity within the observation season (although it is sufficient for calculating the periodicity over an entire observation season).

The importance of planetary-scale waves in the momentum transport around the Venusian cloud-top level, migration of absorbers, and formation of planetary-scale UV brightness features highlights that investigating the long-term continuous changes in these variation can clarify basic wave properties, such as their typical time scales or lifetimes, as well as how these planetary-scale waves control the real Venusian atmosphere. The main scope of this study is observational characterization of the temporal variations in the periodic wind and UV brightness variations.

The successful orbit re-insertion of the Japanese Venus Climate Orbiter (VCO), named Akatsuki, occurred on December 7, 2015 (Nakamura et al., 2016). More than five Venusian years of observations have been conducted by Akatsuki, and the onboard UV Imager (UVI), to date. We analyzed the periodicities in the UV brightness variation over the half of Venusian years by using Akatsuki/UVI images, based on our previous work (Imai et al., 2016), to capture the temporal changes in planetary-scale waves at the Venusian cloud-top level. The cloud-tracked winds were also analyzed using spatially resolved UV images obtained by Akatsuki/UVI to characterize the planetary-scale wave propagation in the fast zonal wind. The rest of this paper is organized as follows. Section 2 describes the details of observations of Akatsuki/UVI and data navigations. Section 3 describes the analytical methodology to capture the temporal changes of waves and presents the results of the periodicity variations in the continuous UV brightness data, which are also compared to the mean zonal wind and periodicity in wind variations. Section 4 discusses the implications of these periodical variations and whether they can be a manifestation of the planetary-scale waves and characterizes these waves. Section 5 summarizes the presented study's conclusions.

## 2 Observations and data navigation

### 2.1 Akatsuki/UVI observations and image processing

Akatsuki/UVI has two observation channels at 283- and 365-nm wavelengths (Yamazaki et al., 2018); we only use the 365-nm images in this study. The 365-nm channel monitors the cloud or brightness of the UV absorbers at the ~70 km cloud top. Akatsuki is orbiting around Venus in the near equatorial plane with ~10.5-day period. UVI obtains Venus images at ~2-hour intervals during a nominal observation sequence. Continuous Venus observations can be conducted when the Akatsuki orbit's apoapsis is located in the dayside region, which is suitable for tracking cloud motions. Here we focused on the prominent periodic and planetary-scale variations in winds and cloud brightness, whose 3–5-day periods were previously reported in Del Genio and Rossow (1990) and Kouyama et al. (2012, 2013, 2015). We investigated the temporal variability in June–October 2017, when clear continuous periodic variations were observed. Since the amplitude of the periodic signals were low and/or the periodicity frequently changed less than one month in many other Akatsuki observation seasons, the waves in the present seasons could be more active than usual.

Table 1. Summary of the data used in this study, which were obtained by the 365-nm monitoring channel on Akatsuki/UVI. Orbit number was counted up since the Venus orbit re-insertion occurred on December 7, 2015. Number of data points indicates the number of images used to measure the brightness and the number of cloud motion vector fields, which derived from image triplets (see Section 2.3). The time coverage of brightness data was slightly shorter than that of the wind data because more strict observation geometry was required for the image selection for brightness data (see Text S1 and S2 in the supporting information).

| Data type | Observation season (orbit number) | Number of data points |
|---|---|---|
| UV brightness | June 15 – October 9, 2017 (r0051–r0062) | 961 images |
| Cloud tracking | June 1 – October 30, 2017 (r0050–r0064) | 1067 vector fields |

Table 1 provides a summary of the data analyzed in this study. We selected Venus images with a sub-spacecraft point that existed between 06:00 and 18:00 local time (LT), which approximately corresponds to a phase angle of <90°, for the UV brightness analysis. Images that either suffered losses in the image field, were erroneously flat-fielded, or had incorrect geometry information, were manually excluded. The analyzed data are the level 2b (L2b) and 3bx (L3bx) products provided by Japan Aerospace Exploration Agency, described in Yamazaki et al. (2018) and Ogohara et al. (2017), respectively. (see the supporting information for all the data list). These data are provided in the datasets, VCO-V-UVI-3-CDR-V1.0 under the Planetary Data System standards (PDS3) and vco_uvi_l3_v1.0, respectively (Murakami et al. 2018a; 2018b). The L2b product consists of the calibrated data, with a physical unit of W $m_{-2}$ $sr_{-1}$ $m_{-1}$. Each image was projected onto a longitude (0°–360°) and latitude (60°S–60°N) map, with 0.25° × 0.25° resolution, based on the L3bx geometry information, which is navigated using a limb-fitting technique (Ogohara et al., 2017). Only the valid longitudinal area, defined as the differential angle between the sub-spacecraft and sub-solar longitudes (<75°), is calculated in this projection procedure. The original image's grid points are interpolated using a fifth-degree polynomials from triangles formed by Delaunay triangulation (see Akima (1996)). Figure 1 shows an example of the UVI image and each processed product.

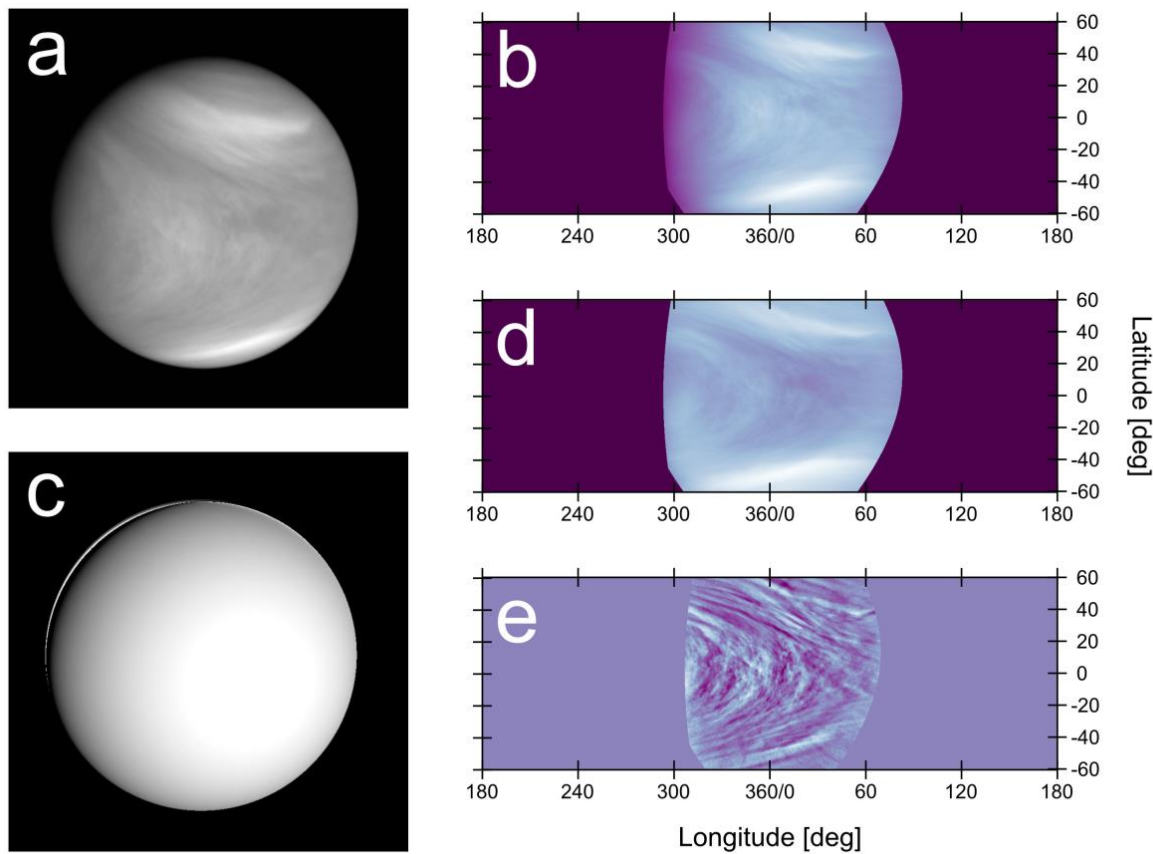

**Figure 1.** Sample 365 nm UVI image obtained 2017-07-30T23:04:74 and obtained products through the image processing. (a) is an original Venus disc image of l2b data and (b) is the projected image onto a longitude (0°–360°) and latitude (60°S–60°N) map. (c) is a reconstructed Venus disc of the scattering properties (see Section 2.2), and (d) is the projected image after the correction of the scattering properties. (e) is a band-pass filtered image as the same manner of Horinouchi et al. (2018) by applying Gaussian filters with half-widths at half-maxima of 4° (high-pass) and 0.3° (low-pass) for both longitude and latitude.

## 2.2 UV brightness data

The 365-nm UV brightness variability within each image, due to the scattering properties of the Venusian clouds, was reduced based on Lee et al. (2017). Their disk function $D$ describes brightness as a function of scattering angle (i.e., emergence ($e$) and incidence ($i$) angles, and the phase angle ($\alpha$)). Following their empirical formula for the 365-nm wavelength, $D$ can be written as the combined Lambert and Lommel-Seeliger law (LLS) (Buratti and Veverka, 1983; McEwen, 1986):

$$D = k(\alpha) \frac{2\mu_0}{\mu_0 + \mu} + \left(1 - k(\alpha)\right)\mu_0, \tag{1}$$

where

$$k(\alpha) = 0.257086 + 0.00102739\alpha - 1.45403 \times 10^{-5}\,\alpha^2, \tag{2}$$

$\mu = cos(e)$, $\mu_0 = cos(i)$, and $k$ is the coefficient describing the relative contributions of Lambert and Lommel-Seeliger, where 0 is the pure Lambert law and 1 is the pure Lommel-Seeliger law. Most of the hot pixels in the original images, caused by cosmic rays, were excluded from the onboard median-combining process of three Venus frames. Dead or bad pixels were treated as negligible in this study.

The absolute brightness changes were simply measured from the LLS corrected data to determine brightness variation from the rotation of planetary-scale UV features. We measured the averaged brightness within a 30° longitudinal range (2-hour LT range) from each image, which is >3000 km at the equator, to enhance these planetary-scale features. Since the sub-spacecraft LT at the apoapsis gradually changed from the morning to evening side, the observable dayside region also gradually shifted. Therefore, we shifted the center LT of the analyzed area based on the apoapsis motion (7 hour per 100 days). The analyzed initial LT coverage was 08:00–10:00 LT on July 1, 2017 (30 day after June 1), and the final coverage was approximately 15:00–17:00 LT. Figure 2 shows the data coverage and the shift of analyzed LT region at 55°S, with these longitudinal coverages illustrating the minimum case since the Akatsuki orbital plane is close to the equatorial plane. We used data from 55°S to 55°N, with 1° resolution, for the latitudinal coverage.

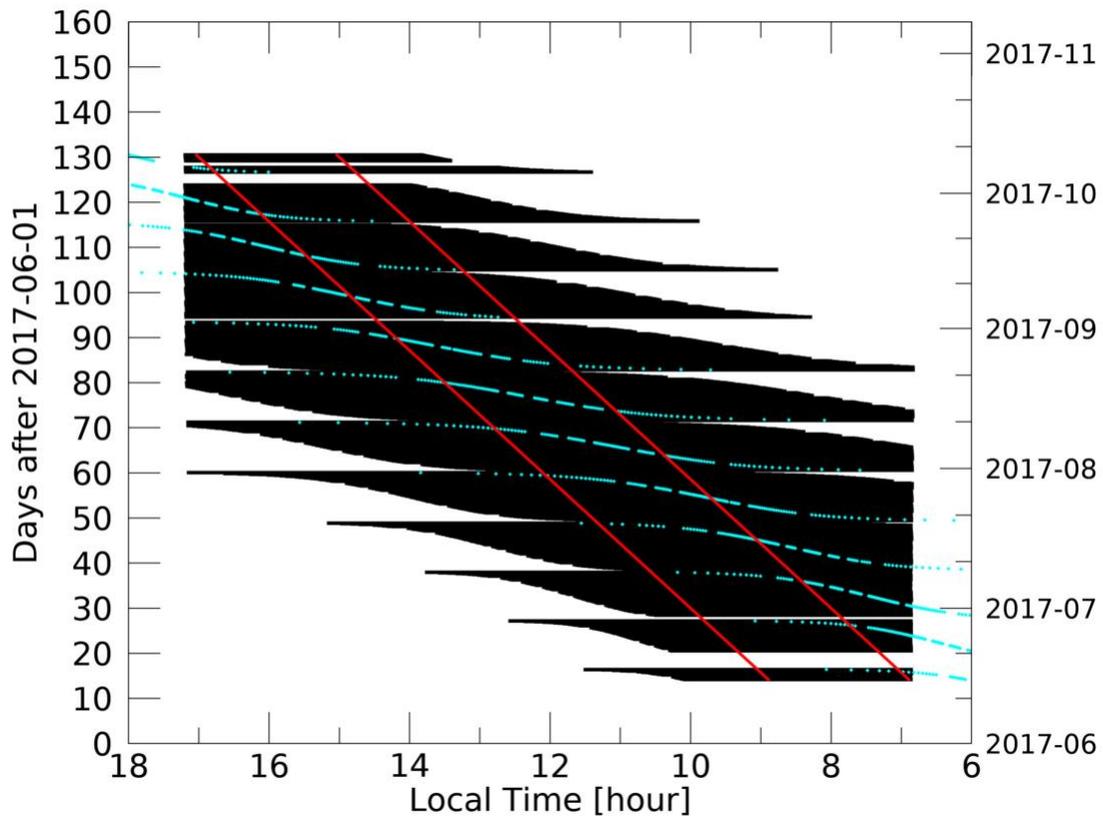

**Figure 2.** Temporal variations in the local time (LT) coverage, indicated by the black-colored region of analyzed UVI images, at 55°S. The cyan dots show the sub-spacecraft LT, and the red lines indicate the LT ranges used in our analysis, which is shifted by 7 hour per 100 days, with the coverage adjusted to 08:00–10:00 LT on July 1, 2017.

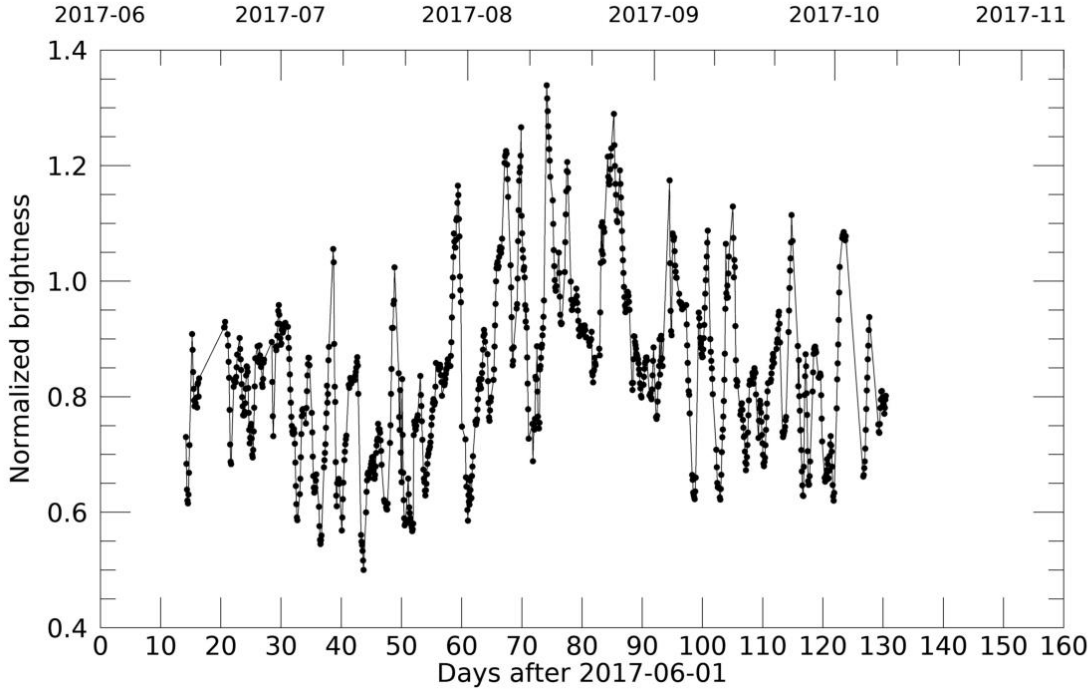

**Figure 3.** An example of the time series of the normalized UV brightness variation in the equatorial region (10°S–10°N).

Figure 3 depicts the measured brightness variations in the equatorial region (10°S–10°N). The photometric property correction, outlined in equations (1) and (2), was first applied to the data, and the UV brightness time series was then normalized by the typical disc mean value. This mean brightness was calculated in the 55°S–55°N latitude range through the entire observation season. There were two periodic brightness variations, with ~4–5- and ~10-day periods, in the data (Figure 3). This latter periodic variation is likely due to the periodic changes in the observation phase and emission angle condition, which is attributed to the observation geometry and the 10.5-day orbit period. We did not attempt to remove this extraneous signal since our target period is considerably shorter than 10 days.

## 2.3 Cloud-tracked wind data

We analyzed the cloud-tracked wind data and investigated the wind variations to clarify the existence of planetary-scale waves. We only provide a brief outline of cloud-tracking here since the detailed cloud-tracking methodology is given by Ikegawa and Horinouchi (2016), Horinouchi et al. (2017), and Horinouchi et al. (2018). A two-dimensional Gaussian high-pass filter with $\sigma = 3°$ was used in this method to emphasize the image contrast, which allowed tracking typical small-scale features (~300 km) at the equator. The peaks in the cross-correlation surface, derived from successive image triplets with 2-hour intervals, were used to retrieve the cloud motion vectors. The observation time for each derived vector field was defined as the average of acquisition time of the input image triplet. On the other hand, the locations of the vectors were approximated by shifting the initial template-region center by 7.5° to the west (or, equivalently, adding 0.5 hours to LT) in order to roughly compensate the overall advection due to the super-rotation (for example, a zonal movement by 100 m/s cause longitudinal displacement of 6.7°). The zonal and meridional winds were measured in each $3° \times 3°$ grid point; some derived wind vectors were removed from the analysis if either the error index (called eps ($\varepsilon$) in Ikegawa and Horinouchi (2016)) was >20 m s−1 or they were inconsistent with neighboring vectors. The error index in both the zonal and meridional directions were measured from the sharpness of the cross-correlation surface. The consistency was checked using the so-called relaxation method (see Horinouchi et al. (2017)). Further parameters used for the cloud-tracking were described in Horinouchi et al. (2018).

Figure 4a and 4c shows the LT-dependent structures of the zonal and meridional winds obtained by averaging the velocities on the LT-latitude coordinates and 4e depicts the number of vectors used for averaging. Here, we applied a smoothing over three grids (9°) in the latitudinal dimension. The LT dependence is mostly attributed to thermal tides; the dependence is similar to those reported in previous works (e.g., Khatuntsev et al., 2013; Hueso et al., 2015). The number of cloud motion vectors is small in the higher latitude because of the low contrast of small-scale features due to the observation geometry and/or streaky cloud morphologies. Figure 4b and 4d represent corresponding standard deviations of zonal and meridional winds respectively. In the equatorial region, $\sigma_u$ ~10 m s-1 and $\sigma_v$ ~5 m s-1, and they are increased about factor of two towards mid-latitudes. These standard deviations should reflect the amplitude of waves, other perturbations, and measurement errors. Here, we use the cloud motion vectors from 55°S to 55°N, spanning the same latitude range as the UV brightness data.

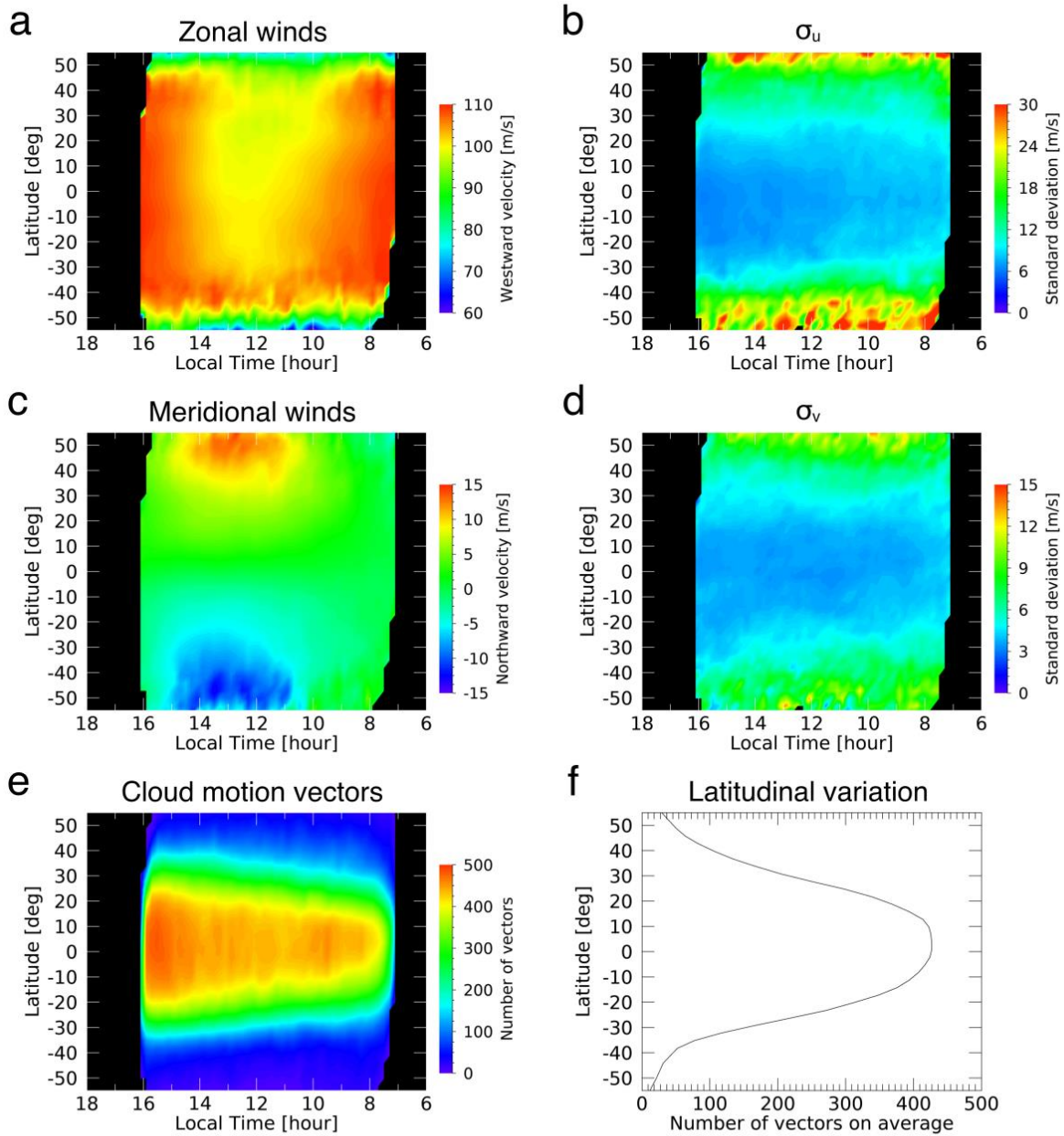

**Figure 4.** LT-dependent structures in (a) the westward velocity and (c) northward velocity, and the corresponding standard deviation of (b) zonal and (d) meridional winds. (e) shows the number of vectors used for averaging and (f) is the line plot of latitudinal variation of the number of vectors on average. Each figure was obtained by smoothing over three grids (9°) with latitude.

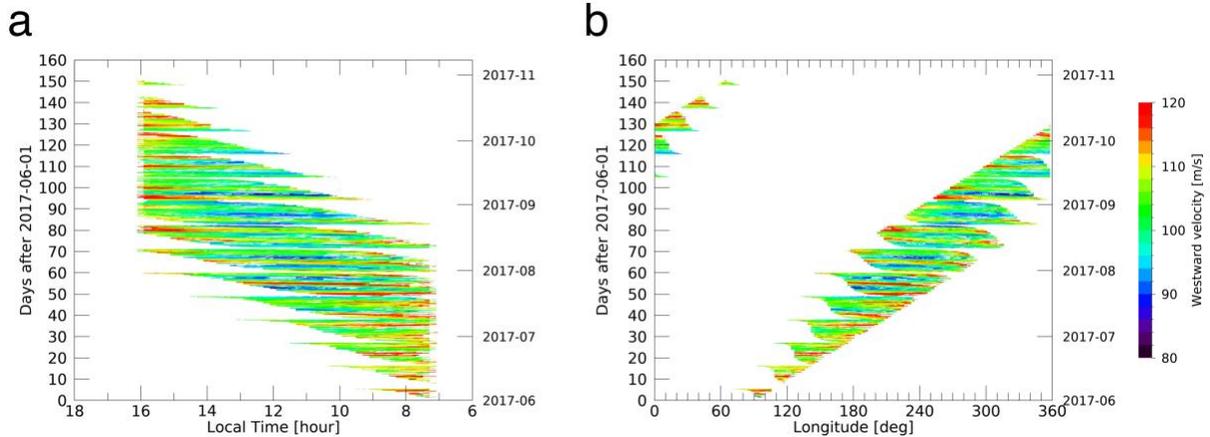

**Figure 5.** Time series of the westward zonal wind variations in the equatorial region (2.25°S ± 1.5°) as functions of (a) LT and observation time (b) longitude and time.

Figure 5a (5b) displays a Hovmöller diagram of the zonal wind at the equator as functions of LT (geographical longitude) and observation time. Clear LT dependence can be confirmed, and some periodic variations superimposed on that structure. On the other hand, we did not find significant topographic dependence on zonal winds. Bertaux et al. (2016) reported a large longitudinal variability in the long-term-mean cloud top winds (with a peak-to-peak difference greater than 20 m/s) by using data from Venus Express/VMC. However, Fig. 5 indicates that the topographic variation is smaller than those of the ~4–5 days periodic variation and the LT dependence (thermal tide); this is the case in an earlier period of Akatsuki's observation (Horinouchi et al, 2018). Further study would be needed to solve the conundrum, but it is beyond the scope of this study. In order to obtain the time series of wind data having no LT dependence, we subtracted the LT-dependent structure from all cloud motion vectors and adjusted the to the dayside (08:00–16:00 LT) mean velocity. Although the LT dependent structure could have some variability and detail analysis is required to model the structure, we simply defined time-averaged wind fields (Figure 4a and 4c) as the LT-dependent structure during the observation season. Figure 6 shows zonally averaged zonal wind in the equatorial region, where the LT dependence was removed. There are obvious short-time-scale (<10 days) periodic variations, which could be attributed to the propagating planetary-scale waves, as reported in previous studies (e.g., Del Genio and Rossow, 1990; Kouyama et al., 2015).

A longer trend in dayside mean zonal wind was also observed, as indicated by the blue line in Figure 6, which possessed an approximately constant value of ~105 m s–1. Since we could not continuously observe the dayside region at the same LT coverage throughout an observation season, it was hard to correctly distinguish a long-term trend in zonal wind variations at time scales of several months from the LT dependence. Our procedure to remove LT dependence may also exert a de-trending effect on the time series at the same time, such that the obtained dayside mean zonal wind may not necessarily be constant. However, since our retrieved LT dependence, whose amplitudes were less than 10 m s–1, was similar to those reported in previous studies (e.g., Hueso et al., 2015; Horinouchi et al., 2018), a larger variation in zonal mean wind (>10 m s–1) might not exist during this observation season.

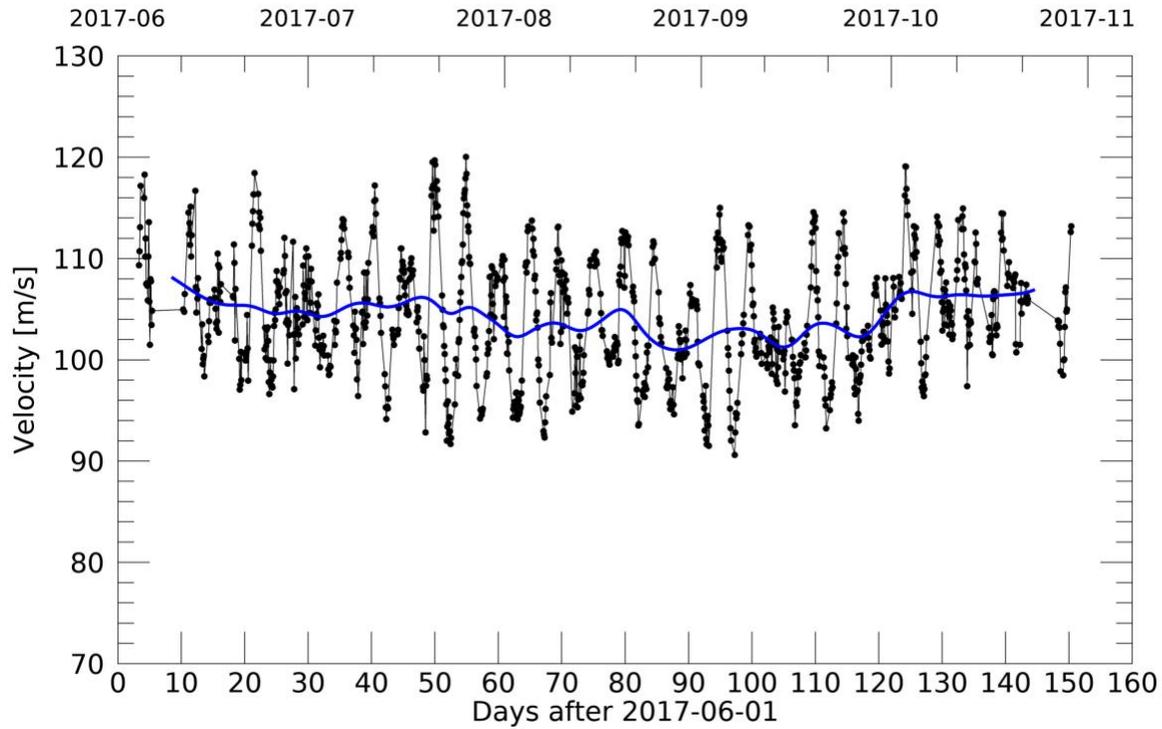

**Figure 6**. An example of the time series of zonal wind in the equatorial region (10°S–10°N) after subtracting the LT dependence. The blue line represents the smoothed curve with a Gaussian kernel with a sigma value of 30 data bins.

## 3 Results

3.1 Total periodicity during the observation season

Periodic wind variations could be a manifestation of planetary-scale atmospheric waves. We first conducted a periodic analysis of the zonal and meridional winds during the entire observation season, verifying which wave types were significant in the focused observation season. Figure 7 shows the amplitude spectra of the zonal and meridional winds as functions of period and latitude. Here, the period ($P$) is measured on the planetocentric coordinate system. Since the same LT region gradually shifts westward over 116.7 day (one Venusian solar day), the corresponding angular velocity of westward movement of a LT region is $\omega_{LT} = 2\pi/116.7$, and $P = 2\pi/(\omega_{app} - \omega_{LT})$, where $\omega_{app}$ is the apparent angular velocity derived from the observed periodicity in time series data. The white line denotes the corresponding period of the dayside mean zonal wind, which is shown by the blue line in Figure 6, as a function of latitude, where 100 m s–1 roughly corresponds to the 4.3-day period at the equator. The periodicities were obtained using a Lomb-Scargle periodogram (Lomb, 1976; Scargle, 1982; Horne and Baliunas, 1986). This analysis can extract the power spectra (square of the amplitude) from unevenly sampled data, which is equivalent to fitting sinusoidal variations to the data via a least-squares method. The periodogram, which estimates the power spectrum (square of the amplitude) and false alarm probability (FAP), are obtained during the calculation to assess the significance of a given periodic signal in the data (see Appendix A).

The 5.1-day period, the dominant mode during this observation season, was slower than the dayside mean zonal wind for each latitude region. The zonal wind variation was dominant in low latitudes of 30°S–35°N, and the meridional wind variation had large amplitudes in the mid-latitudes (~45°). On average over the observation season, 5.1-day mode had the amplitude of ~5.5 m s-1 in both zonal and meridional winds. The peaks of 3.5–4.0-day modes, which were faster that the dayside mean zonal wind, were concentrated in the equatorial region. Although these modes have rather weak amplitude of ~2 m s-1, the calculated 99% of the significance level derived from the Lomb-Scargle analysis indicates that the amplitude larger than ~1.5 m s-1 in zonal winds at the equator can be distinguished from noise and other variations. There were some fragment modes with >99% significance, but these amplitudes were close to the 99% significance level.

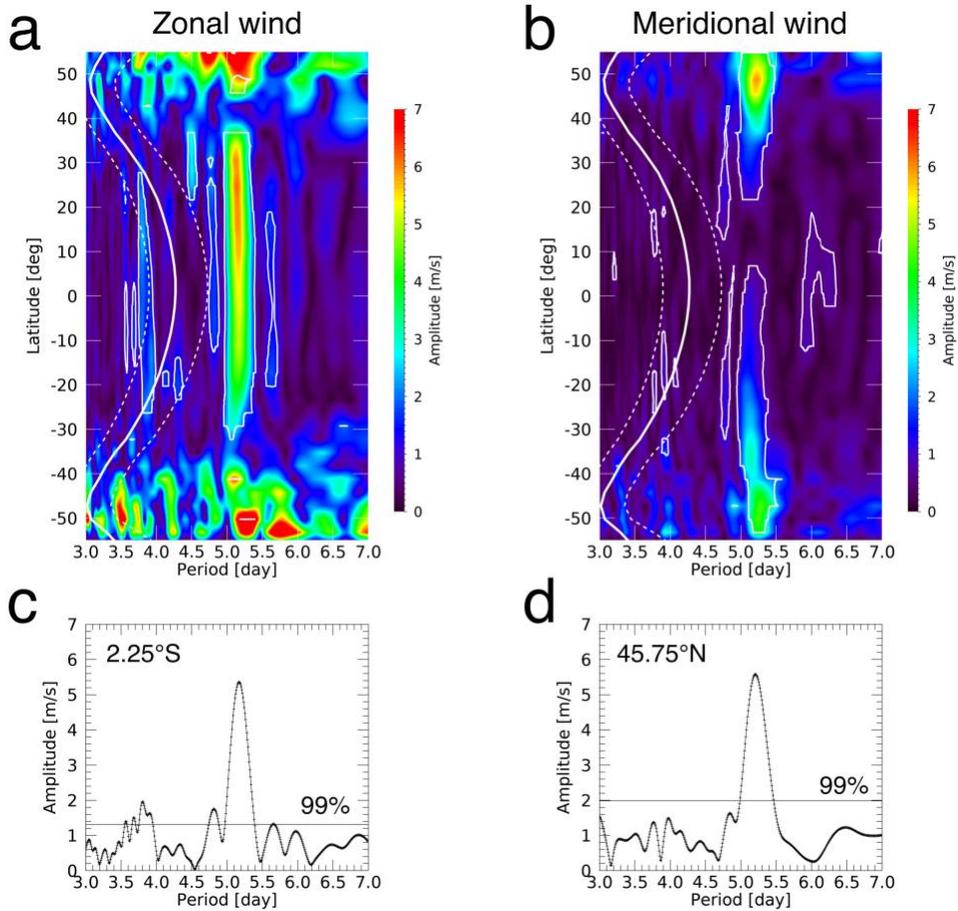

**Figure 7.** Amplitude spectra of the zonal (a) and meridional (b) wind as a function of latitude obtained from the Lomb-Scargle periodogram analysis. The colored regions enclosed by the thin solid white lines possess >99% significance. The thick solid white line and dashed white lines indicate the corresponding period of the background dayside mean zonal wind and the ±10 m s–1 statistical variation, respectively. Panel (c) and (d) represent the amplitude spectra of the zonal wind at 2.25°S and meridional wind at 45.75°N, and horizontal lines indicate the 99% significance level.

### 3.2 Temporal variations in periodicity and amplitude

We previously suggested that the periodicity in UV brightness variations was subject to temporal variations at time scales of a few months (Imai et al., 2016). Therefore, we conducted a time-shifting Lomb-Scargle periodogram analysis (hereafter called the TSLS periodogram analysis) to successive time blocks extracted from the original time series. Figure 8 depicts the schematic illustration of the TSLS periodogram analysis. We set the length of each time block to $L$, with a center time $t'$, and computed the Lomb-Scargle periodogram for the data $X(t)$ at time $t$, where $t' - L/2 < t < t' + L/2$; $t'$ was then incremented by 8 hours. The spectral peaks became sharper when we increased $N$ for the data subset, which gave the spectral peaks a large enough significance against the noise. However, the temporal resolution, which was estimated as half the time range $L$ (see Appendix B), decreased for large $L$, and large $L$ also reduced the total analyzable time range in the observation season. We chose $L=$ 30 day after exploring a series of time ranges; this value provided a good balance between the temporal resolution and observation season coverage. Since this time range was only ~6 times longer than the 5-day period, the obtained spectra were broadened in the period direction, making it difficult to determine the exact peak period. However, we could confirm two modes, with 3.5- and 5.0-day periods, that were distinguishable in the data, and concluded that 30 day was appropriate for detecting the periodic signals in our dataset, providing sufficient temporal resolution and total coverage of the observation season. The ability to detect the change in periodicity is demonstrated using three test model datasets (Appendix B).

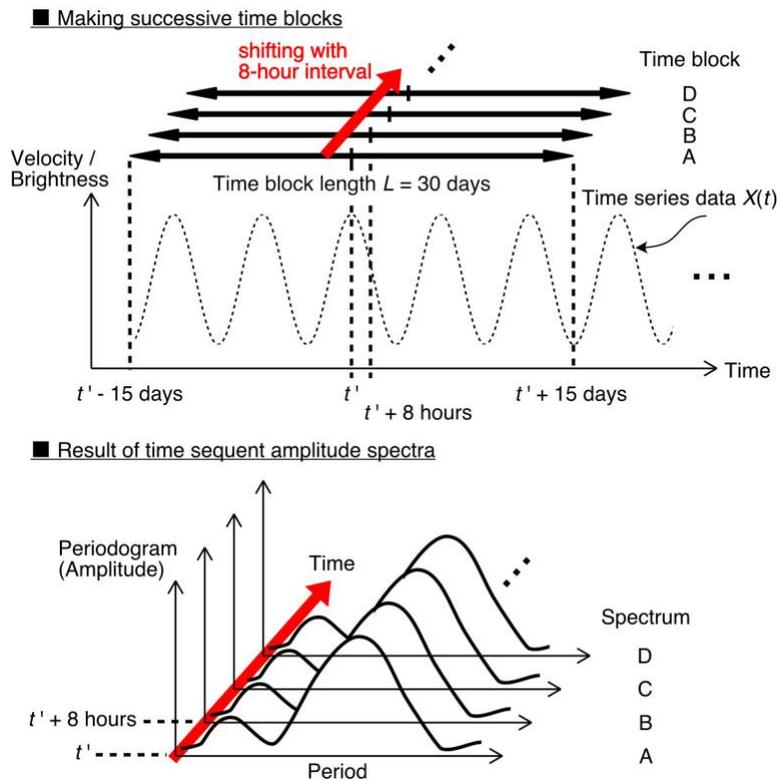

**Figure 8.** Schematic illustration of the TLSL periodogram analysis capturing the temporal changes in amplitude spectra from the time series of winds and brightness.

The constant shift of the center LT of the analyzed area (7 hour per 100 days, as introduced in Section 2.2) used during the TSLS periodogram analysis of the UV brightness data was also accounted for in determining the periodicity, with $P = 2\pi/(\omega_{\mathrm{app}} - \omega_{\mathrm{LT}} + \omega_{\mathrm{shift}})$, where $\omega_{\mathrm{shift}} = 2\pi \frac{7\,\mathrm{h}}{24\,\mathrm{h}} / 100$. Furthermore, we confirmed no clear LT dependence on the periodicity of the brightness variations. Figure 9 shows the temporal changes of the spectra of winds and brightness. Each of the nine panel is the result of TSLS periodogram analysis, and three different latitudinal regions of 30°S–50°S, 10°S–10°N, and 30°N–50°N were investigated separately. Similar to Figure 7, we observed periodicities in the 3–7-day range, with the white line denoting the corresponding period of the dayside mean zonal wind for each latitude band.

The ~5-day period was a prominent mode in the UV brightness results of both mid-latitude regions and the equatorial region at the >99% significance. The corresponding signal appeared in both the zonal wind variation in the equatorial region and meridional wind variation in mid-latitude regions. The peak amplitudes of the equatorial zonal wind and mid-latitude meridional wind variations for the 5-day mode were ~8 and ~5 m s–1, respectively, reaching their maximum variations ~50–60 day ($t = 50$–60 day) after the reference date of June 1, 2017 ($t = 0$ day). However, the confirmed peaks in mid-latitude brightness variations for the same 5-day mode became significant at $t = 70$–80 day, indicating the enhanced periodic wind variation delayed the brightness variation peaks in the UV brightness data. A ~5-day mode was present in the equatorial brightness variations, but their amplitudes were relatively small and possessed cyclic amplitude variations.

The observed prominent 5-day modes suggest Rossby waves, as previously reported in Del Genio and Rossow (1990) and Kouyama et al. (2015). Furthermore, a continuous, non-negligible 3.8-day mode was confirmed in the UV brightness and zonal wind variation in the equatorial region, whereas there was no clear 3.8-day variation in the meridional wind. Although the amplitude and significance of the 3.8-day mode are relatively low, this mode should correspond to a wave that propagates faster than the background super-rotation, such as a Kelvin wave. We will discuss the characteristics of these two wave modes in the following section and Section 4.

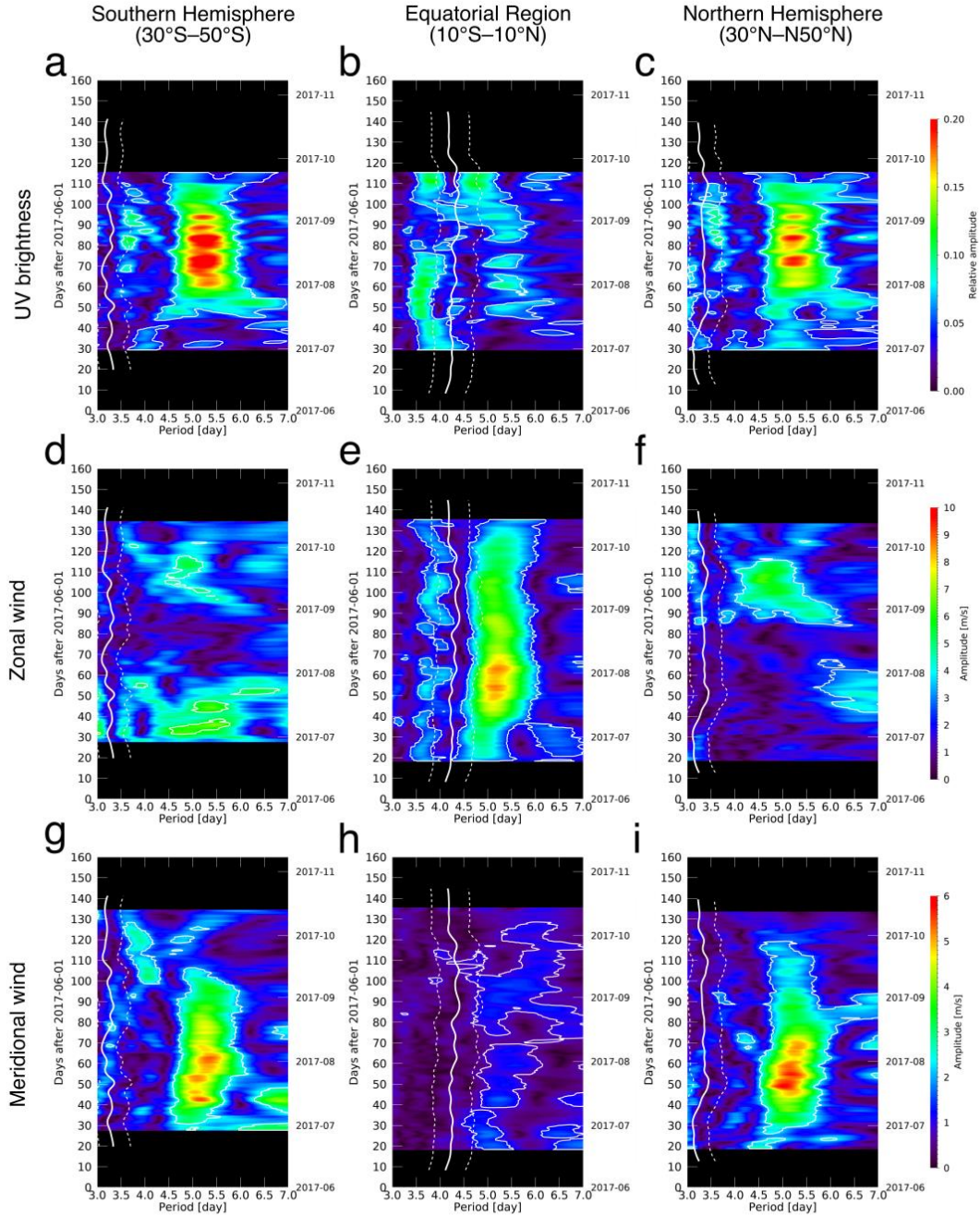

**Figure 9.** Temporal changes in amplitude spectra of the UV brightness (a–c), and zonal (d–f) and meridional (g–i) wind as a function of observation time for each latitude region: 30°S–50°S (left), 10°S–10°N (middle), and 30°N–50°N (right). The colored regions enclosed with thin white lines denote the regions with >99% significance. The thick solid white line and dashed white lines indicate the corresponding period of the background dayside mean zonal wind and ±10 m s₋₁ uncertainty, respectively, which were calculated in the same manner as the blue line in Figure 6.

The 5-day mode existed during most of the observation season and exhibited a temporal amplitude evolution. Therefore, we divided the observation season into three representative sub-seasons, "pre-peak" ($t = 17$–$47$ day; June 18 to July 18), "peak" ($t = 47$–$77$ day; July 18 to August 17), and "post-peak" ($t = 77$–$107$ day; August 17 to September 16), which were based on the zonal and meridional wind amplitudes, to investigate their temporal changes. Figure 10 shows the amplitude spectra of the UV brightness, and zonal and meridional winds as a function of latitude for each sub-season, which spanned the ~08:00–10:00, ~10:00–12:00, and ~12:00–14:00 LT regions, respectively. Here we used a data length of 30 days, which is the same as the window size for the TSLS periodogram analysis.

The UV brightness variations appear to exhibit both 3.5–4.0- and 5.0–5.5-day modes, as shown in Figure 10. While the 3.5–4.0-day mode was dominant in the equatorial region (25°S–25°N) and possessed an attenuating trend, the 5.0–5.5-day mode was dominant in the mid-latitude regions (>35°) and possessed increasing amplitudes. The zonal wind data possessed a relatively weak 3.5–4.0-day mode in the equatorial region, which was observed in each sub-season. However, the 5.0–5.5-day mode was apparently subjected to temporal amplitude variations. The 5.0–5.5-day mode existed in the meridional wind data, but there was no significant 3.5–4.0-day mode. The representative periodicities in the UV brightness, and zonal and meridional wind data, are summarized in Table 2. Here we should note that the amplitudes of the zonal and meridional winds, retrieved at higher latitudes (>45°), were erroneously high for most of the spectral range. Venus often has streaky and white obscured features in the mid- to high-latitudes. This might explain the high uncertainty and sparseness of the original cloud-tracking data due to the initial screening (based on the uncertainty of the measured value). The elongation of these high-amplitude streaks in the zonal direction suggests that the error was particularly severe in the zonal component. Details on the dispersion and sparseness of the used cloud-tracking vectors are provided in Appendix C.

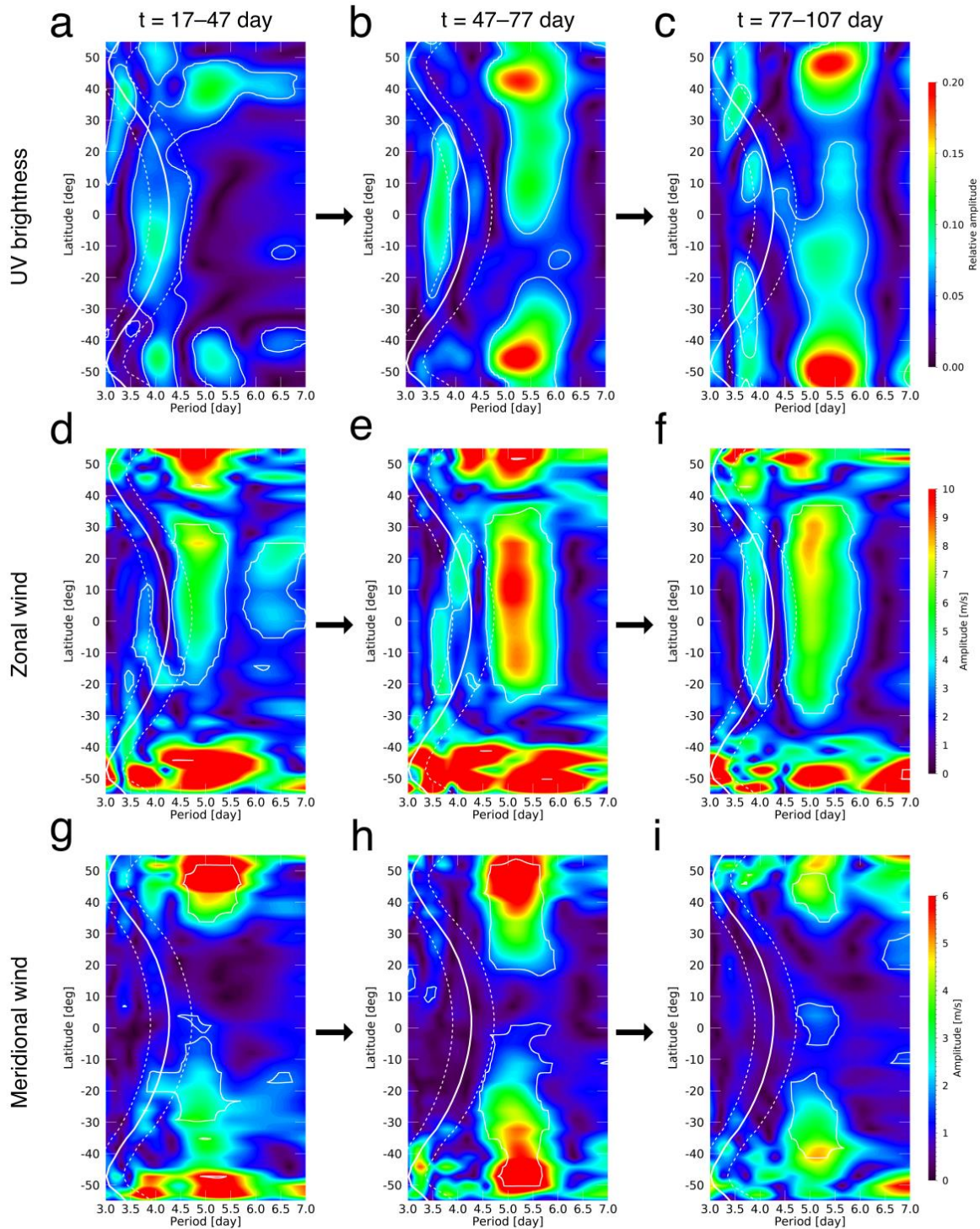

**Figure 10.** Amplitude spectra of the UV brightness (a–c), and zonal (d–f) and meridional (g–i) wind as a function of latitude. Each sub-season is presented here: "pre-peak" ($t = 17$–$47$ day; June 18 to July 18), "peak" ($t = 47$–$77$ day; July 18 to August 17), and "post-peak" ($t = 77$–$107$ day; August 17 to September 16).

**Table 2.** Periodic characteristics of the UV brightness, and zonal and meridional winds for the equatorial and mid-latitude regions. The significant modes, which possess either Kelvin-like (3.5–4.3 days) or Rossby-like (4.5–5.5 days) periodicities, are denoted by K and R, respectively. The periods shown in brackets represent the ranges at the >99% significance. The corresponding propagation velocity $c$ was calculated from the bold wind period, and the intrinsic phase velocity $c - \bar{u}$ was obtained using the dayside mean zonal wind ($\bar{u}$) for each latitude region.

| Observation sub-season | | UV brightness period (day) | Zonal wind period (day) | Meridional wind period (day) | propagation velocity $c$ (m s$^{-1}$) | Dayside mean zonal wind $\bar{u}$ (m s$^{-1}$) | Intrinsic phase velocity $c - \bar{u}$ (m s$^{-1}$) |
|---|---|---|---|---|---|---|---|
| Equatorial region (2.25°S ± 1.5°) | | | | | | | |
| "pre-peak" | (K) | 3.8 (3.4–4.2) | 3.7 (3.6–4.0) | — | 120.3 | | 16.1 |
| ($t$ = 17–47 day) | (R) | — | 4.8 (4.5–5.4) | 5.0 (4.8–5.2) | 92.7 | | −11.5 |
| "peak" | (K) | 3.5 (3.3–3.8) | 3.6 (3.5–3.9) | — | 123.6 | 104.2 | 19.4 |
| (t = 47–77 day) | (R) | 5.2 (4.8–5.5) | 5.1 (4.6–5.5) | 5.1 (4.6–5.5) | 87.2 | (4.3 days) | −17.0 |
| "post-peak" | (K) | — | 3.8 (3.7–4.1) | — | 117.1 | | 12.9 |
| ($t$ = 77–107 day) | (R) | 5.3 (4.5–5.5) | 5.1 (4.5–5.5) | — | 87.2 | | −17.0 |
| Mid-latitude region (45.75°N ± 1.5°) | | | | | | | |
| "pre-peak" | (K) | — | — | — | — | | |
| ($t$ = 17–47 day) | (R) | 4.8 (4.5–5.5) | — | 5.1 (4.7–5.5) | 63.8 | | −37.6 |
| "peak" | (K) | — | — | — | — | 101.4 | |
| ($t$ = 47–77 day) | (R) | 5.0 (4.5–5.5) | — | 5.1 (4.6–5.5) | 63.8 | (3.2 days) | −37.6 |
| "post-peak" | (K) | — | — | – | — | | |
| ($t$ = 77–107 day) | (R) | 5.1 (4.6–5.5) | — | 5.0 (4.7–5.4) | 65.1 | | −36.3 |

### 3.3 Structure and temporal variations of the Rossby-like wave

Since the propagation velocity of the observed prominent 5-day mode was slower than that of the background zonal wind, it could be a planetary-scale Rossby wave, with retrograde propagation that is opposite to the background super-rotation. We obtained the wave's amplitude and phase by fitting a sinusoidal curve to the zonal and meridional winds at each latitude from the Lomb-Scargle periodogram calculations. Therefore, we reconstructed the wind field's horizontal structure from only the observed 5-day mode; here, we assumed the observed periodic signal as a wave with zonal wavenumber 1.

Figure 11 depicts the reconstructed horizontal wind fields for the three sub-seasons, "pre-peak" ($t = 17$–$47$ day; June 18 to July 18), "peak" ($t = 47$–$77$ day; July 18 to August 17), and "post-peak" ($t = 77$–$107$ day; August 17 to September 16), where we selected the peak period of the Rossby-like wave in each sub-season as 5.1, 5.1, and 5.0 day, respectively. These three periods are considered as the same 5-day mode. The same LT ranges, ~08:00–10:00, ~10:00–12:00, and ~12:00–14:00 LT, were used for the reconstructions. The relative vorticities $\xi_{(i,j)}$ [s-1] at each longitude and latitude coordinate $(\lambda_i, \theta_j)$ were shown in the figure as color contours, which were calculated as follows:

$$\xi_{(i,j)} = \frac{v_{i+1} - v_{i-1}}{a \cos \theta_j (\lambda_{i+1} - \lambda_{i-1})} - \frac{u_{j+1} \cos \theta_{j+1} - u_{j-1} \cos \theta_{j-1}}{a \cos \theta_j (\theta_{j+1} - \theta_{j-1})}, \qquad (3)$$

where $u$ and $v$ [m s-1] are the eastward zonal and northward meridional winds, respectively, and $a$ is the planetary radius ($a = 6052 \pm 70$ km for Venus). The suffixes $i = 0, 1, \ldots, 118$ and $j = 0, \ldots, 36$ indicate the grid position, with $\lambda_i = (3i + 3)/180 \times \pi$ and $\theta_j = (3j - 54)/180 \times \pi$ [rad]. We note that the large absolute vorticity at higher latitudes could be erroneous due to the large cloud-tracking uncertainty. However, we have confirmed that similar wind structures were retrieved when we used different LT ranges in the reconstruction (for example, we changed the range from 11:21–13:21 LT to 08:00–10:00 LT for the peak sub-season). The reconstructed wind field should be feasible since the different LT datasets have different characteristics.

The prominent planetary-scale vortices, which were symmetric around the equator, can be confirmed during the peak sub-season (middle panel of Figure 11). The typical relative vorticity was $1 \times 10$-5 s-1 at the center of the vortex at 40° (both hemispheres), which is the same order as the planetary vorticities of the super-rotation. The maxima of the westward and eastward zonal winds in the equatorial region exhibits a $\pi/4$ phase difference compared to the poleward and equatorward meridional wind peaks in the longitudinal direction. Since Akatsuki is located approximately above the equatorial region in its orbit, its observation condition may be suitable for extracting the equatorially symmetric structure, which has never been confirmed before. The vortices were not clear during the pre-peak sub-season's initial stage (30 day before the peak sub-season), especially in the northern hemisphere. However, weak symmetrical vortices existed in the post-peak sub-season (30 day after the peak sub-season).

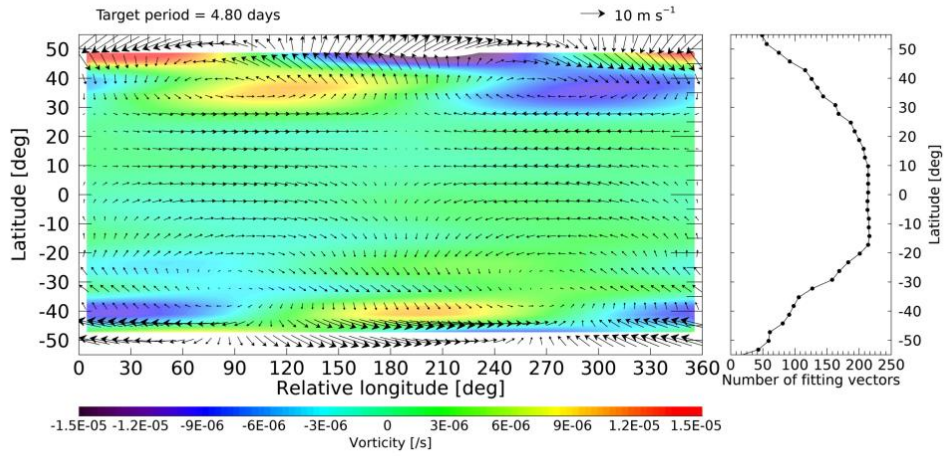

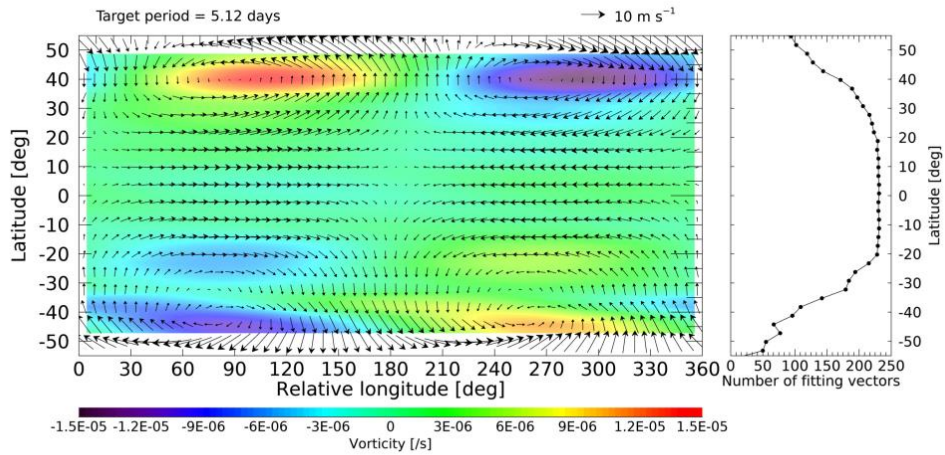

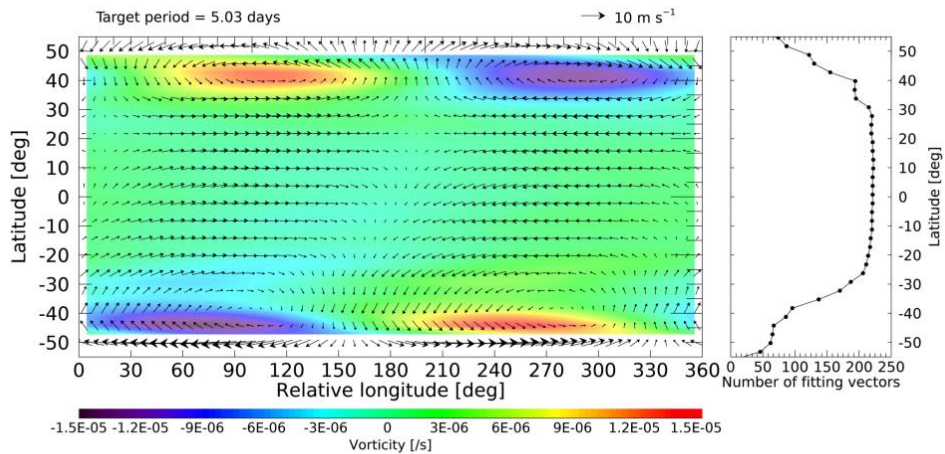

**Figure 11.** Horizontal wind vectors targeting the Rossby-like wave, with a ~5-day period, which were reconstructed using the amplitude and phase derived from the Lomb-Scargle periodogram analysis for each sub-season: $t = 17–47$ (a), $47–77$ (b), and $107–137$ day (c). The target periods, shown at the top right of each figure, were used to calculate the amplitude and phase. The reconstructed wind fields were smoothed with $3 \times 3$ grids ($9°$ longitude $\times 9°$ latitude) and shifted in the longitudinal direction as the zero-phase zonal wind at the equator, which was equal to 0 at the relative longitude, for a better visual comparison between the sub-seasons. The wind field changes from up to down, with a 30-day time interval between each panel. Right plots depict zonally averaged latitudinal variations of number of wind vectors used for the fitting. While the number of vectors decreases to less than 100 in higher latitudes, most of the 5.1-day modes in both zonal and meridional winds have > 99% of significance as shown in Figure 10. The color contours represent the relative vorticity, and we excluded the higher latitudes (>45°) due to the large uncertainties in the absolute wind speeds and low number of fitted vectors.

## 3.4 Temporal variations in the planetary-scale UV brightness features

We can expect the variation in the UV brightness since the observed Kelvin-like and Rossby-like waves, and their evolutions, could alter the horizontal distribution of UV absorbers (e.g., Yamamoto and Takahashi, 2012; Nara et al., 2019). Figure 12 shows the temporal changes in the planetary-scale UV brightness for each sub-season. UVI images projected onto a longitude-latitude map with $0.25° \times 0.25°$ resolution were composed assuming a 4.3-day circulation, which corresponds to the equatorial dayside mean zonal wind of 104 m/s. There are some unnatural brightness discontinuities, which may reflect the imperfect geometry-dependence model for the brightness.

We observed the periodic appearance of dark clusters during the pre-peak sub-season in the equatorial region as denoted with red lower triangles in Figure 12. The dark clusters should be the main source of the observed ~4-day periodicity in this sub-season, as shown in leftmost part of the top panel of Figure 10. In the peak sub-season, the periodic dark clusters gradually disappeared, and the white belts in the 45°–60° latitude regions in both hemispheres became rippled periodically in the latitude direction, as shown by the red arrows in Figure 12. While the rippled belts were more prominent in the southern hemisphere, we observed that they changed synchronously between the northern and southern hemispheres. These rippled white belts were still present in the post-peak sub-season, but they gradually disappeared, and any prominent planetary-scale features were not sustained. The equatorial planetary-scale UV brightness features underwent dramatic changes throughout the entire observation season.

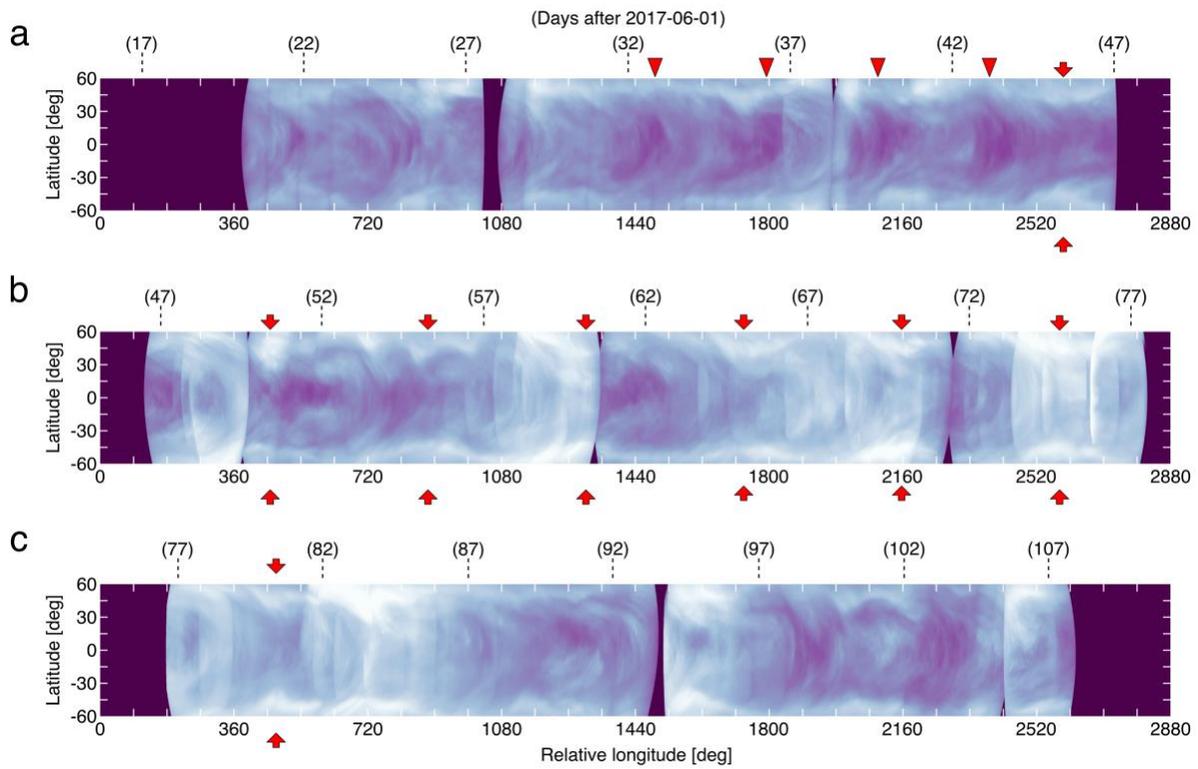

**Figure 12.** Composite UVI images projected onto relative longitude and latitude (60°S–60°N) map, with 0.25° × 0.25° resolution. Each composite image was constructed by using images obtained in each sub-season $t = 17$–47 (a), 47–77 (b), and 107–137 day (c), and each brightest pixel was used in overlap regions. Numbers in the parentheses show the approximate days after June 1, 2017 when UVI observed the LT regions of 08:00–10:00, 10:00–12:00, and 12:00–14:00 respectively. The red lower triangles mark the dark clusters in the equatorial region, and the red arrows denote the prominent rippled white belts that extend to the lower latitude region.

## 4 Discussion

### 4.1 Amplification of the 5-day Rossby-like wave

The Rossby-like wave structure was the dominant mode during the observation season. Kouyama et al. (2013) found planetary-scale vortices, with centers located at 40°S–45°S, in the VMC cloud-tracking wind data. We identified vortices in the same latitude region in both hemispheres, revealing an equatorially symmetric structure. This symmetric structure could be a manifestation of the equatorial mode of the Rossby wave, which has been reported in numerical studies (e.g., Imamura et al., 2006), and hereafter we regard the 5-day wave as the Rossby wave. In the previous studies e.g. Yamamoto and Takahashi (2003), Sugimoto et al (2014), and Lebonnois et al (2016), the structures of planetary-scale waves and their influences on the super-rotation were investigated by several Venusian atmospheric general circulation models. More recently, the symmetrical planetary-scale streak structures were discovered in the middle- and lower-cloud altitudes using the 2-μm camera (IR2) on board Akatsuki (Peralta et al., 2018; 2019), and Kashimura et al. (2019) demonstrated the importance of equatorial Rossby-like waves in the formation of the observed planetary-scale streak-like structures in AFES-Venus, a dynamical Venusian atmospheric general circulation model. Here, we investigate the characteristics of the ~5-day wave under that assumption that the observed 5-day variation is the equatorial mode of the Rossby wave.

Figure 13 depicts the temporal changes of the meridional velocity amplitude associated with the 5-day wave in the northern mid-latitude region (30°N–50°N; red line), where the meridional wind variation was significant. A clear amplification started at $t = 30$ day and persisted for ~20 days at a rate of ~0.18 m s$_{-1}$ day$_{-1}$. On the other hand, it took ~50 days for the amplitude to return to the initial level.

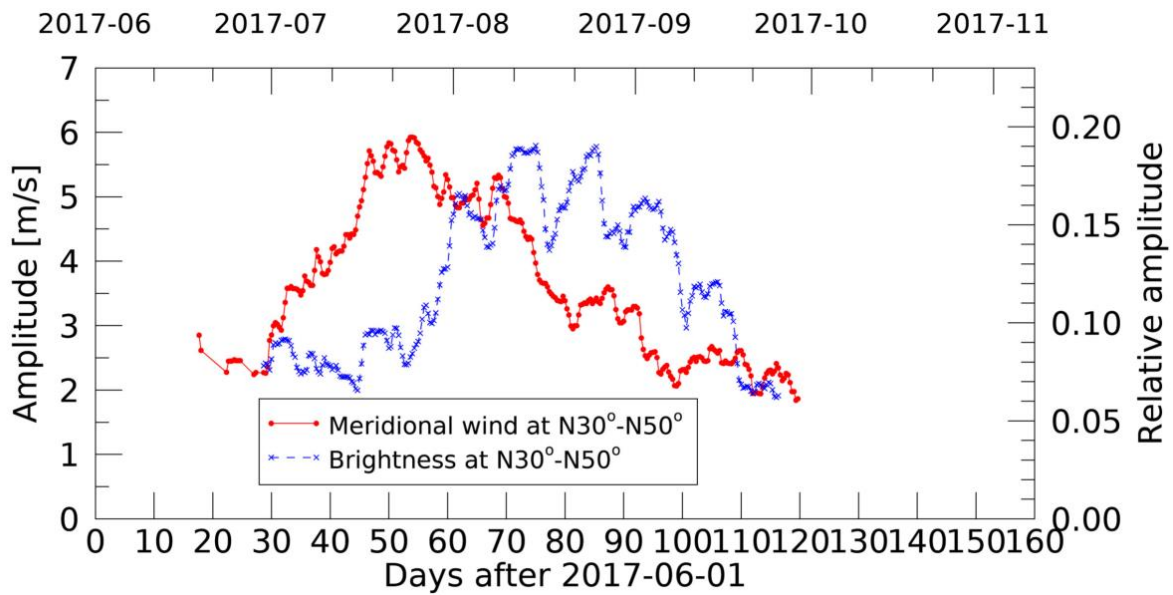

**Figure 13.** Temporal changes in the amplitude of the meridional wind (red dots) and UV brightness variations (blue crosses) in the northern mid-latitude region (30°N–50°N). The amplitudes were calculated via the TSLS periodogram analysis, with the peak sub-season possessing 4.5–5.5-day periods at >99% significance.

The observed time scale of the dramatic Rossby wave amplitude variations might be related to the temporal variation in the wave source and/or the vertical wave propagation. If the wave propagates from the source region to the cloud top in a short time period, then the observed ~20 day of amplification and ~50 day of attenuation can be attributed to changes in the wave source. Although the observed Rossby wave had a broad latitude structure, we assumed that the wave's basic characteristics could be discussed using equatorial wave theory due to its equatorially symmetric structure. The dispersion relation for the low-frequency equatorial Rossby wave can be written as (cf. Chapter 4 in Andrews et al., 1987):

$$m = N\beta\omega^{-2}\left\{\left(n+\tfrac{1}{2}\right) - \left[\left(n+\tfrac{1}{2}\right)^2 + \omega k\beta^{-1}(1+\omega k\beta^{-1})\right]^{\frac{1}{2}}\right\}, \tag{4}$$

under the beta-plane approximation for simplicity, where $k$ and $m$ are the horizontal and vertical wavenumbers, respectively, $n$ is the meridional mode number (or the order number of Hermite function in meridional direction) and we set the equatorial mode to $n = 1$. $N$ is the Brunt-Väisälä frequency, $\omega$ is the intrinsic frequency, and $\beta$ is the meridional derivative of the Coriolis parameter. We calculated the intrinsic frequency as $\omega = (c - \bar{u})k$, where we used the intrinsic phase velocity $(c - \bar{u})$ in Table 2 and $k = 1/a$. Furthermore, we calculated the gradient of planetary vorticities as $\beta = 2\Omega\cos\theta_0/a$, with $\theta_0 = 0°$, and set the atmosphere's rotation rate in the equatorial region to $\Omega = 2\pi(4.3\text{ day})^{-1}$ (Table 2). We obtained $\beta = 5.5 \times 10^{-12}$ m$^{-1}$ s$^{-1}$, $m = 3.6 \times 10^{-4}$ m$^{-1}$, and a vertical wavelength of $\lambda_z = 2\pi/m = 17$ km for $N \sim 2.0 \times 10^{-2}$ s$^{-1}$ and $c - \bar{u} = -17$ m s$^{-1}$ (Table 2).

The vertical group velocity can then be calculated as:

$$c_g^{(z)} = \frac{\partial\omega}{\partial m} = -\frac{\omega^3}{N\beta}\left\{2\left[\frac{3}{2} - \sqrt{\frac{9}{4} + \frac{\omega k}{\beta}\left(1+\frac{\omega k}{\beta}\right)}\right] + \frac{1}{2}\frac{\frac{\omega k}{\beta}\left(1+2\frac{\omega k}{\beta}\right)}{\sqrt{\frac{9}{4} + \frac{\omega k}{\beta}\left(1+\frac{\omega k}{\beta}\right)}}\right\}^{-1}. \tag{5}$$

We obtained $c_g^{(z)} = 2.0 \times 10^{-2}$ m s$^{-1}$ (1.7 km d$^{-1}$), which corresponds to ~3 day to travel a scale height of $H \sim 5$ km. Radiative cooling should be the most important mechanism for damping the wave at the cloud top. The relaxation time scale of the radiative damping is ~6–10 day at the cloud top for the observed disturbances with a vertical wavelength of 17 km based on the estimation of Crisp (1989). Therefore, the Rossby wave should propagate vertically without strong wave attenuation.

Temporal changes in the amplitude of UV brightness variation are also confirmed (Figure 13). We observed a clear correlation between the amplitude changes of the meridional wind and UV brightness at a time scale of ~2 months. Since the planetary-scale vortices shown in Figure 11 can latitudinally transport absorbers, the observed Rossby wave, whose vortex centers are located in the mid-latitude regions, can be responsible for the UV brightness variations. For example, the rippling of white belts in the mid-latitudes could be due to the meridional transport of clouds and absorbers via vortex flow. These planetary-scale UV brightness variations can be interpreted as being directly induced by the Rossby wave vortices.

A ~20-day time lag is seen between the amplitude changes in the wind and the brightness: the wind amplitude increased first and the increase in the brightness amplitude followed. We have confirmed a similar 20-day lag also for a reduced latitude range of 30°N–40°N. We could not find any clear reason for this time lag, however, the rippling white belts in Figure 12 is observed since at least $t = 45$ day, when the meridional wind variation reached its maximum of ~5 m s$_{-1}$. The appeared 20 days lag might be related to the time scale for the 5-day periodical brightness change, but there is also the possibility of an artifact.

## 4.2 Short-period zonal wind and UV brightness variations associated with the Kelvin-like wave

The UV brightness variation in the equatorial region exhibited a significant 3.8-day periodicity, as well as the ~5-day mode. Since the dayside mean zonal wind of 104 m s$_{-1}$ was slower than the observed 3.8-day periodic rotation speed (equal to 120.3 m s$_{-1}$ rotation at the equator), the existence of a prograde propagating Kelvin wave might explain the >15 m s$_{-1}$ velocity difference. Although the brightness variations alone cannot confirm the existence of a Kelvin wave, the corresponding wind variation should be expected if the UV brightness is caused by a Kelvin wave (hereafter we simply called as the Kelvin wave). Kelvin waves essentially have atmospheric oscillations in the zonal and vertical directions at lower latitudes. The observed 3.8-day wind variation was only confirmed in the zonal component. It was restricted to the low-latitude regions during the observations, which are consistent with Kelvin wave characteristics, suggesting the existence of a Kelvin wave. Figure 14 shows the observed in time series of the zonal wind and the brightness in the equatorial region (10°S–10°N). Although the 5.1-day mode was dominant in the zonal wind, we can still trace the 3.8-day variation with help of the 3.8-day fitting curve. Detection of the same periodicity from both zonal wind and brightness can support the existence of the Kelvin wave. However, while the calculated zonal wind amplitude of 3.0 m s$_{-1}$ was larger than the 99% significance level of 2.3 m s$_{-1}$, the fitting RMS of 3.7 m s$_{-1}$ was also relatively large. The ~5-day Rossby-wave was dominant in the observation season and additional 6.3-day of unknown variation was superimposed to the Kelvin-wave signals. Therefore, it should be noted that the 3.8-day signal captured in this observation season was marginal.

If we assume that the atmospheric super-rotation replaces the planetary rotation, we can predict the Kelvin wave's intrinsic phase velocity based on a linearized wave theory and using an amplitude profile of the zonal wind variation, which should decrease with latitude as a Gaussian function. However, the Kelvin wave's latitudinal structure could not be determined from our periodic analysis due to the insufficient significance of the 3.8-day mode.

Here we characterize the observed 3.8-day wave as a Kelvin wave. When the vertical wind shear is negligible and a radiative damping is absent, the vertical wavelength of Kelvin wave can be estimated as (cf. Chapter 4 in Andrews et al., 1987):

$$\lambda_z = 2\pi \left( \frac{N^2}{(c - \bar{u})^2} - \frac{1}{4H^2} \right)^{-1/2}, \tag{6}$$

where $g$ is the gravitational acceleration, $c$ and $\bar{u}$ are the phase velocity of the Kelvin wave and dayside mean zonal wind, respectively, and $c - \bar{u}$ is the intrinsic phase velocity of the Kelvin wave. We calculated $\lambda_z \sim 5.1$ km for $N = 2.0 \times 10_{-2}$ s$_{-1}$, the scale height $H \sim 5$ km, and $c - \bar{u} = 16$ m s$_{-1}$ (Table 2) assuming the dayside mean zonal wind was similar to the mean zonal wind, which was measured from both the dayside and nightside regions. Del Genio and Rossow (1990) estimated $\lambda_z \sim 6$ km, which is consistent with our calculation, and not far from that obtained by Kouyama et al. (2012) ($\lambda_z = 9 \pm 3$ km).

Although the detail of the vertical distribution of UV absorbers is unknown, several in situ photometric measurements by Venera-13, -14 and VEGA1, 2 indicates absorbers exist at altitudes lower than the cloud deck (Ekonomov et al., 1983; Bertaux et al.1996). If the observed 3.8-day variation reflects a Kelvin-wave, the dark clusters concentrated in the equatorial region (as shown in Figure 12) should be interpreted as the upwelling regions, where the absorber was supplied from lower levels due to vertical material transport via the Kelvin wave. Here we assume that a) the Kelvin wave propagates in a vertical upward direction, b) the chemical lifetimes of the absorbers are longer than the 3.8-day period, and c) the sensing altitudes of the brightness and cloud-tracking wind data are the same, or more specifically, the small-scale UV features used for cloud-tracking are located at the altitude where the optical thickness is $\tau \sim 1$. Under these assumptions, the phase of the darkest region in the brightness data should be delayed by 90° relative to the point of maximum zonal velocity (e.g., Fig. 9 in Del Genio and Rossow, 1990) since the absorbers are lifted upward over a half cycle of the wave. However, according to the observation (Figure 14) in the pre-peak sub-season, when the 3.8-day mode was confirmed in both the UV brightness and zonal wind data, the darkest region was close to, or even slightly earlier than, the maximum of the zonal velocity; that is inconsistent with the above expectation. The same phase relationship has already been reported in Del Genio and Rossow (1990) and Kouyama et al. (2012). Del Genio and Rossow (1990) found that the darkest region lagged the maximum zonal velocity by 0 or 1 day when the 4-day wave was active. Kouyama et al. (2012) also found a clear in-phase correlation, with the dark (bright) region corresponding to the fast (slow) westward zonal wind region. Therefore, these three results indicate an approximately in-phase correlation.

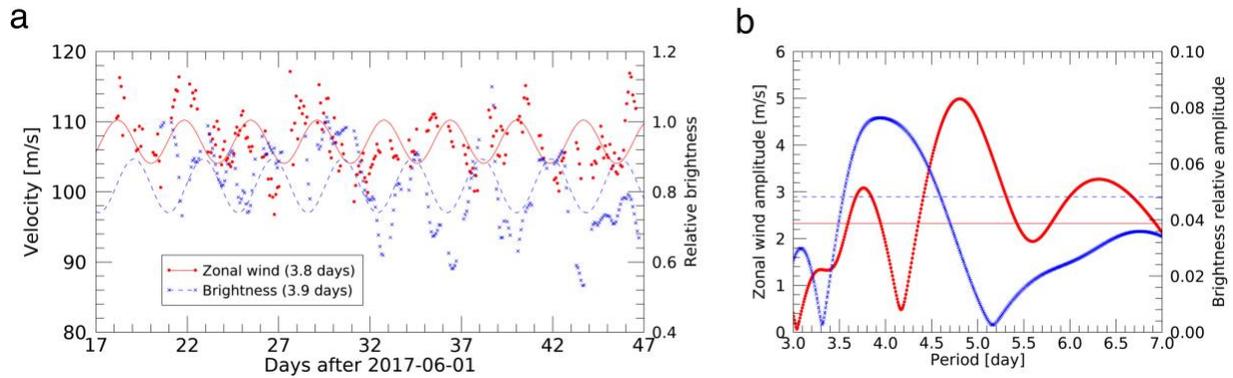

**Figure 14.** (a) Time series of the westward zonal wind (red dots) and UV brightness (blue crosses) variations obtained in the equatorial region (10°S–10°N) for ~08:00–10:00 LT in the pre-peak sub-season. The plotted lines were best-fit sinusoidal curves having 3.8- and 3.9-day periodicities for the wind and brightness data respectively, and their RMS were 3.7 m s-1 and 0.099. In the plot, the dominant 5.1-day signal was extracted from the original zonal wind. (b) shows the amplitude spectrum of the original data. The amplitude of zonal wind (brightness) for the 3.8-day mode was 3.0 m s-1 (0.08), and the 99% significance level was 2.3 m s-1 (0.05) indicated by the horizontal line.

We assumed that the Kelvin wave propagates upward in assumption a), whereas a downward-propagating Kelvin wave would produce the opposite phase relation. Since upwelling occurs in the slower (eastward) zonal wind field in the downward propagation case, this downward propagation cannot explain the observed almost zero phases. A review of assumption b) indicates the zero-phase relation could be explained by the absorbers' shorter lifetime, where the darkest region occurred at the strongest upwelling wind. This scenario was proposed in Kouyama et al. (2012). The difference in the altitude between the UV brightness and the cloud-tracked velocity may also explain the phase relationship even if the lifetime of the absorber is long. Since the scale of the small features (~300 km) used for cloud-tracking are largely different from that of planetary-scale features (>5000 km), we assume there is a significant sensing altitude difference. Using the typical vertical wavelength of a Kelvin wave (6 km), the observed in-phase relation between the westward zonal wind and darkness in the UV brightness data could be explained by an altitude difference of $\Delta h = -4.5$ km (1.5 km) for a UV brightness sensing altitude that is at a lower (higher) altitude than that of the wind measurements, as shown in Figure 15.

Del Genio and Rossow (1990) mentioned that the altitude of large-scale UV contrasts could be below the cloud top level of $\tau \sim 1$ (= 70 km), which was estimated via polarimetry analysis, and above about $\tau \sim 5$ (65–66 km). They argued that, if either the cloud-top level or cloud-tracking sensing altitude was further than approximately one-half of $\lambda_z$ from the sensing altitude of large-scale UV features, then their observed phase difference should be reasonable. Our suggestion of $\Delta h = -4.5$ km follows their indications. However, since the UV absorbers' vertical distribution is largely unknown, and other factors may influence the observed temporal changes, further investigations are necessary to elucidate the origin of these planetary-scale UV features and the role of Kelvin waves in the formation of the dark regions.

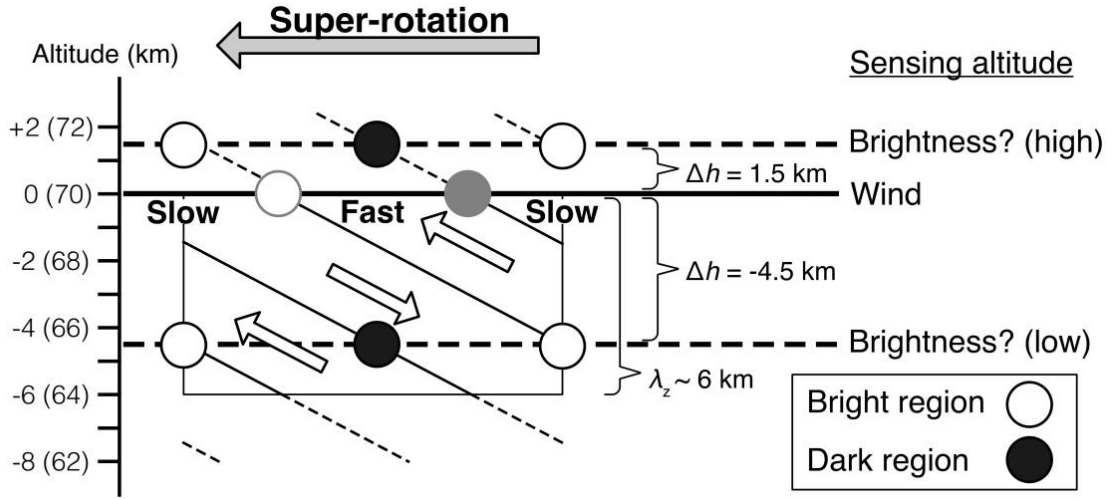

**Figure 15.** Schematic illustration of the longitude-height section along the equator showing an upward-propagating Kelvin wave's vertical structure and the relationship between the observed zonal wind and UV brightness. The reference altitude of 70 km (as shown in parentheses) is set to the sensing altitude of the wind data, and the white and black filled circles denote the bright and dark regions, respectively. The gray open and filled circles indicate the expected brightness distribution, but they are inconsistent with previous and current observations. Introducing either $\Delta h = -4.5$ or 1.5 km differences in the sensing altitudes between the wind data (70 km) and brightness data yields an in-phase relationship between the fast zonal wind and the darkest UV region.

4.3 Temporal evolution of apparent planetary-scale UV features

The dominant wave modes seemed to change from a Kelvin wave to a Rossby wave throughout the entire observation season in this study, with attenuation of the Rossby wave observed at the end of the observation season. These temporal variations can easily be inferred from the observed spatiotemporal evolution of the UV features. Our results suggest the Kelvin wave could be important to form dark clusters in the equatorial region. The Rossby wave can contribute to alter the brightness distribution as rippling the latitudinal contrasts and have large impacts in mid-latitude regions, and dark features in the equatorial region became unstable. The absence of prominent waves should be the reason of no sustainable planetary-scale features.

By the analysis of Venus Express VMC images, Nara et al. (2019) recently demonstrated the formation of the Y-shaped feature with a simplified transport model including both Kelvin and Rossby wave effects. In their scenario, the dark materials are supplied to the cloud top level in the equatorial region by a Kelvin wave, and poleward advection by the meridional circulation and Rossby wave vortices combined with zonal stretching by the mid-latitude jet play important roles. Although we could not identify the Y-shaped features due to their large variability, similar brightness features could be confirmed in mid-latitude regions during the peak sub-season. They also reported the UV features in the absence of the Rossby wave. Their result showed a dark cluster around the equator with the northeast-southwest (southeast-northwest) tilted structures in the southern (northern) hemisphere. They also showed the latitudinal shear of the zonal wind can contribute to form slightly tilted structures, and the same characteristics can be confirmed during the pre-peak sub-season when the Kelvin-wave alone was active (Figure 12a). Peralta et al. (2015) investigated the temporal changes of the UV features from the Kelvin wave and distortions of its structure by winds, however, our result showed the almost fixed pattern over the four cycle (> 16 days). Therefore, the Rossby-wave or interaction between Kelvin and Rossby wave could be the reason of large temporal change in planetary-scale UV features.

## 5 Conclusions

We investigated the continuous planetary-scale wave variations in the Venusian atmosphere using recent UV brightness and wind data obtained by Akatsuki/UVI and revealed that the planetary-scale waves underwent a temporal evolution and degradation at monthly time scales. We selected the June–October 2017 observation season, which is when two significant wave modes were observed. The first mode was the 5.1-day Rossby wave; its zonal wind variation in the equatorial region, and meridional wind in the mid-latitude regions, exhibited remarkable amplitude change accompanied by prominent variations in UV brightness.

The reconstructed horizontal structure of the 5.1-day mode revealed planetary-scale vortices in both hemispheres with equatorial symmetry. These vortices caused the meridional transportation of UV absorbers and clouds, with white belts in the 45°–60° latitude regions of both hemispheres synchronously formed rippled features in the latitudinal direction. The observed Rossby wave's relatively long vertical wavelength suggests the expected radiative damping time scale is shorter than the propagation time scale. The observed ~20 day of amplification, or ~50 day of attenuation, in the wind amplitude could be interpreted as the time scale of the wave source change.

The appearance of the Rossby wave is preceded by 3.8-day variation in both the UV brightness and zonal wind data in the equatorial region. While the amplitude and significance of the 3.8-day mode were not as large as the observed Rossby wave in this time, a candidate for this variation is the equatorial Kelvin wave. The vertical upwelling is a potential reason for the dark features in the equatorial region. The observed in-phase relationship between the westward zonal wind and dark regions in the UV brightness probably indicates the difference in sensing altitude of the zonal wind and planetary-scale brightness data.

The dominant wave mode switched from the Kelvin wave to the Rossby wave during the observation season, with attenuation of the Rossby wave observed at the end of the observation season. These temporal variations in wave mode have an apparent effect on deforming planetary-scale UV features (or the famous Y-feature), and dramatically alter the cloud morphology and horizontal albedo distribution at the cloud-top level.

Kouyama et al. (2015) argued that the Kelvin and Rossby waves' vertical momentum transport provides an important clue for the super-rotation's acceleration and deceleration. Therefore, we could expect the temporal variations in planetary-scale waves to have an impact on Venusian atmospheric dynamics, from a momentum transport between different altitudes. The dayside mean zonal wind was almost constant, at ~105 m s$_{-1}$, with no obvious temporal trend. However, our subtraction procedure of the LT dependence may impose a de-trending effect, which implies the obtained dayside mean zonal wind may not necessarily be constant. While this point should be carefully investigated, one reason could be the difference between the sensing altitudes of the wind and brightness data. Analyzing a future observation season, when the mean zonal wind either exhibit large variation or are relatively slower compared to those in the current study, may provide the necessary avenue to elucidate the connection between planetary-scale waves and the super-rotation.

The main mechanisms of Kelvin and Rossby wave excitation are still largely unknown. The observed Rossby wave became amplified immediately after the Kelvin wave disappeared. The Rossby wave had a clear, equatorially symmetric structure, with a synchronized amplifying vortex in both hemispheres, as shown in Figure 9. Such observations may provide important inferences on the wave excitation mechanisms. Future studies should include numerical modeling of the observed temporal evolution of planetary-scale waves and the long-term statistical characterization of planetary-scale waves to elucidate the temporal evolution of Kelvin and Rossby waves in the Venusian atmosphere.

## Acknowledgments

This study was based on the Akatsuki/UVI observational data provided by ISAS. We are grateful for support from all the members of the Akatsuki mission. All original Akatsuki/UVI data are available via Data ARchives and Transmission System (DARTS), provided by Center for Science-satellite Operation and Data Archive (C-SODA) at ISAS/JAXA, and NASA PDS Atmospheres Node. The analyzed data are the L2b and L3bx products, and these data are provided in the datasets VCO-V-UVI-3-CDR-V1.0 (https://doi.org/10.17597/ISAS.DARTS/VCO-00003) under the PDS3 and vco_uvi_l3_v1.0 (https://doi.org/10.17597/ISAS.DARTS/VCO-00016) respectively. (See the supporting information for all the data list). The derived time series of wind data and data plotted in figures can be found in the Mendeley data (http://dx.doi.org/10.17632/kvpxhwvfj5.2). This work was supported by the Japan Society for the Promotion of Science (JSPS) (Grant-in-Aid for JSPS Research Fellow: JP17J03862 and KAKENHI: 16H02225, 19K14789). The authors thank Dr. Ricardo Hueso and an anonymous reviewer for a thoughtful and detailed review of this paper.

## Appendix A: Lomb-Scargle periodogram analysis

The Lomb-Scargle method (Scargle, 1982) calculates the spectral power $P$ at an angular frequency $\omega$ from evenly or non-evenly spaced data points $X_j = X(t_j)$, $j = 1, \ldots, N$ as:

$$P(\omega) = \frac{1}{2} \left\{ \frac{[\sum_j X_j \cos \omega(t_j - \tau)]}{\sum_j X_j \cos^2 \omega(t_j - \tau)} + \frac{[\sum_j X_j \sin \omega(t_j - \tau)]}{\sum_j X_j \sin^2 \omega(t_j - \tau)} \right\}^2,$$

where $X_j$ is subtracted by its arithmetic average ($1/N \cdot \sum_i X_i$), and $t_j$ is the time. $\tau$ is defined by the relation:

$$\tan(2\omega\tau) = \frac{\sum_j \sin 2\omega t_j}{\sum_j \cos 2\omega t_j}.$$

This procedure is equivalent to a linear least-squares fit of the following $X'(t)$ function to the target data $X$, where $a$ and $b$ are obtained via the following equations.

$$X'(t) = a \cos\{\omega(t - \tau)\} + b \sin\{\omega(t - \tau)\}$$

$$a = \frac{\sum_j X_j \cos \omega(t_j - \tau)}{\{\sum_j \cos^2 \omega(t_j - \tau)\}^{1/2}}$$

$$b = \frac{\sum_j X_j \sin \omega(t_j - \tau)}{\{\sum_j \sin^2 \omega(t_j - \tau)\}^{1/2}}$$

The amplitude $A$ and the phase $\varphi$ at $t = 0$ are then given by:

$$A = \sqrt{a^2 + b^2}$$

and

$$\varphi = -\omega\tau + \tan^{-1}\left(\frac{a}{b}\right),$$

which allows us to obtain best-fit curve as:

$$X'(t) = A \sin\left\{\omega(t - \tau) + \tan^{-1}\left(\frac{a}{b}\right)\right\}.$$

Assuming the signal $X_j$ is purely noise, then $P$ exhibits an exponential distribution and we can define the probability distribution as (Horne and Baliunas, 1986):

$$p_Z(z)\mathrm{d}z = \Pr\{z < Z < z + \mathrm{d}z\} = \exp(-z)\mathrm{d}z,$$

where $Z = P_X(\omega)$. When we assume noise with variance unity, then the cumulative distribution function $F$ is:

$$F_z(z) = \Pr\{Z < z\} = \int_0^z p_Z(z')\mathrm{d}z'$$
$$= 1 - \exp(-z).$$

Here, the quantity $\Pr\{Z > z\} = \exp(-z)$, which gives the statistical significance of a large observation power at a preselected frequency, is obtained. This quantity indicates that, as the observed power becomes larger, it becomes exponentially unlikely that such a power level can be due to a random noise variation. We introduce the empirically generated value $N_i = -6.362 + 1.193\,N_0 + 0.00098\,N_0{}^2$ (number of independent frequencies), instead of $N_0$ (number of data points) from Horne and Baliunas (1986), which allows the FAP to be calculated as:

$$\mathrm{FAP} = [1 - \exp(-z)]^{N_i}.$$

This FAP gives the probability that a peak with height $z$ or higher will occur, assuming that the data are pure noise, such that the quantity $1 - \mathrm{FAP}$ gives the probability that the data contain a signal. Here we convert $1 - \mathrm{FAP}$ into a percentage, which we define as the significance.

**Appendix B: TSLS periodogram analysis**

The capability of detecting the periodicity changes, via the Time-Shifting Lomb-Scargle periodogram analysis (TSLS periodogram analysis), as described in Section 3.2, is demonstrated using three test models. These models used two types of periodic signals having data values of $10 \pm 5$ (average base value = 10; amplitude = 5), with 3.5- and 5.0-day wave periodicities (waves A and B, respectively). These waves are combined according to the following formula:

(Simulated wave) = (3.5-day wave) $\times WF_{3.5\text{day}}$ + (5.0-day wave) $\times WF_{5.0\text{day}}$, where $WF_{3.5\text{day}}$ and $WF_{5.0\text{day}}$ represent the weighting functions used to modulate the amplitudes of waves A and B, respectively, and the time intervals between the data points (the sampling points of the simulated wave) of both waves are set to $1 \pm 1.5/24$ day, where a 1.5-hour random variation is applied. Figure 16 shows the three weighting functions for the two waves. Model A demonstrates the immediate change in wave periodicity from 3.5 to 5.0 day at $t = 60$ day. Model B simulates the gradual changes in both waves' amplitudes. Both waves have a constant mean value of 10, with a monotonic decrease in the relative amplitude of wave A (the weighting value in Figure 16) from 5.0 to 0, where $45 \leq t \leq 75$, whereas the amplitude of wave B monotonically increases from 0 to 5.0 in an opposite fashion. Model C includes a monotonic increase and decrease, as in Model B, but the two waves co-exist for 30 day in the middle of the simulated time range. Figure 17 shows the combined wave evolutions for the three models (A–C). The used data points do not have an even time interval of 1 day since we assume scattered data sampling, whereas adding a pure random variation of $\pm 1.5$ hour to the observation time yields a more even time interval, as shown by the open circles in Figure 17.

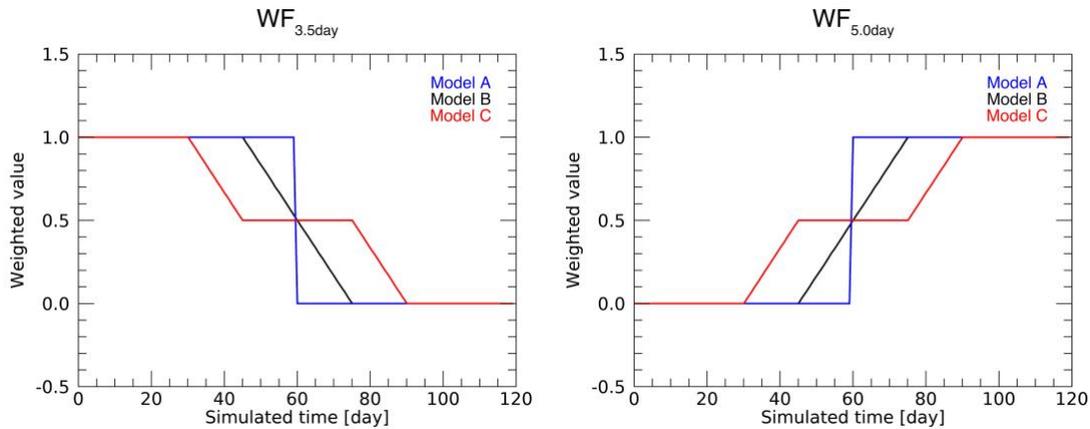

**Figure 16.** Weighting functions for demonstrating the wave change detection capability based on the TSLS periodogram analysis. The time series of the weighting values for the 3.5- (left) and 5.0-day period waves (right) were prepared for Models A–C.

Figure 18 provides the TSLS periodogram analysis results for each model. The horizontal axis is the period, ranging from 3.0 to 7.0 days, and the vertical axis is the observation (simulation) time. The initial and final 15 day contain no TSLS results (black regions in Figure 18) since we chose $\pm 15$ day as the time range of sub-dataset. The colors indicate the amplitude from the Lomb-Scargle analysis (linear scale), and the white dashed lines indicate the >99% significance, which was calculated as $1 - \text{FAP}$. The difference between these three figures helps interpret our results in the following section. Model A shows a clear disappearance of the 3.5-day mode and the appearance of the 5.0-day mode. The periodicity changes at around $t = 60$ day is obscured due to the ±15-day time range. Therefore, we should take the effect of the finite time range into account. The amplitude of the 3.5-day mode in Model B gradually decreases compared to that in Model A, but it is still at the >99% significance, whereas the amplitude of the 5.0-day mode slowly increases from $t = 56$ day. Model C depicts the co-existing state of both modes from $t = \sim$45–75 day, with relatively low amplitude values obtained during this time period. Furthermore, the absolute periods have uncertainties of ~0.2 day for the 3.5-day mode and ~0.4 day for the 5.0-day mode, even if the input signal has a pure periodic variation. However, these uncertainties are low enough to distinguish the two different modes.

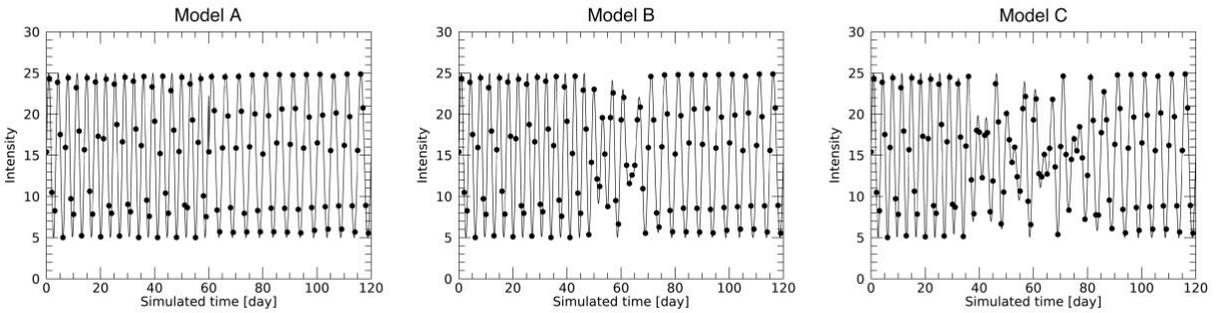

**Figure 17.** Imitation data for demonstrating the detection of wave switching based on the TSLS periodogram analysis for Models A (left), B (middle), and C (right). The open circles denote the ~1-day time interval in each model.

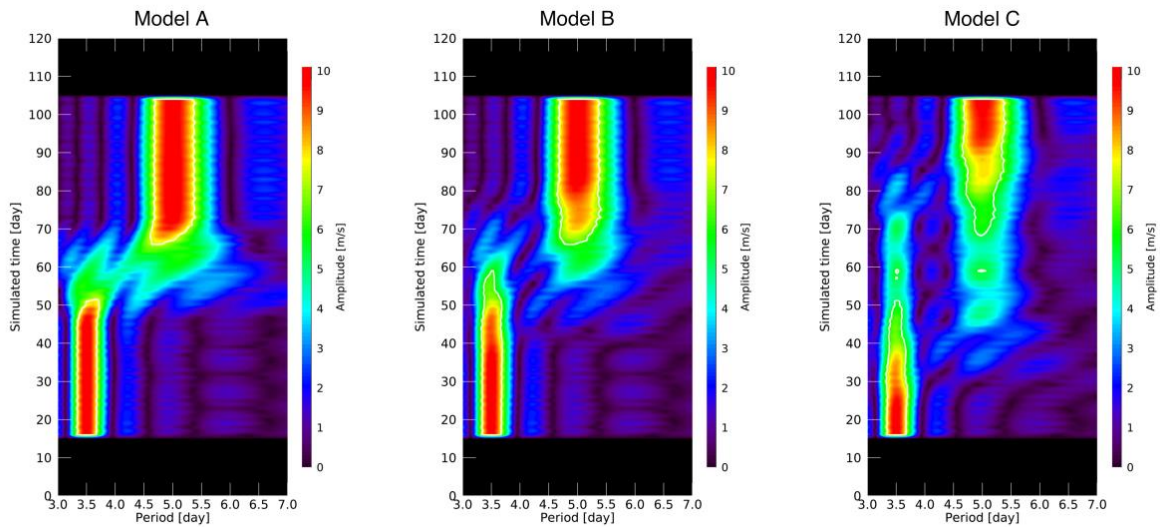

**Figure 18.** Test results of the TSLS periodogram analysis for each model. The colors represent the retrieved amplitudes, and the thin white lines denote the >99% significance. These results demonstrate that the wave and the periodicities changes can be distinguished via TSLS analysis.

**Appendix C: Fitting accuracy for reconstructing the Rossby wave vortices**

We conducted the Lomb-Scargle periodogram analysis of the zonal and meridional winds time series in each latitude region, at 3° intervals, to reconstruct the planetary-scale vortices that accompanied the Rossby wave. Here we show the plausibility of the reconstructed Rossby vortices using the peak observation sub-season ($t = 47$–$77$ day). Figure 19 shows the best-fit results of the calculated sinusoidal curve, which are derived from the amplitude and phase obtained via the Lomb-Scargle analysis. The zonal wind plots are scattered in the mid-latitude region ($>35°$S), resulting in a large fitting error. Since the center of the vortices is located around this latitude, this scatter could be due to the lack of strong wave-induced variation in the zonal direction. While the meridional wind amplitudes are small (~5 m s–1), they are confirmed over a broad latitude range, and the fitting curves seem appropriate, with the small root mean square (RMS) as shown in the top of each panel of Figure 19. The number of data points decreases in the high-latitude regions ($>45°$), due to either the large epsilon value (Ikegawa and Horinouchi et al., 2016) or increased measurement errors of the original cloud-tracking data. Therefore, the derived amplitudes were anomalously high, with no $>99\%$ significance obtained. However, the retrieved phase was unexpectedly consistent with the low-latitude data, which indicates that further careful screening would provide some improvements.

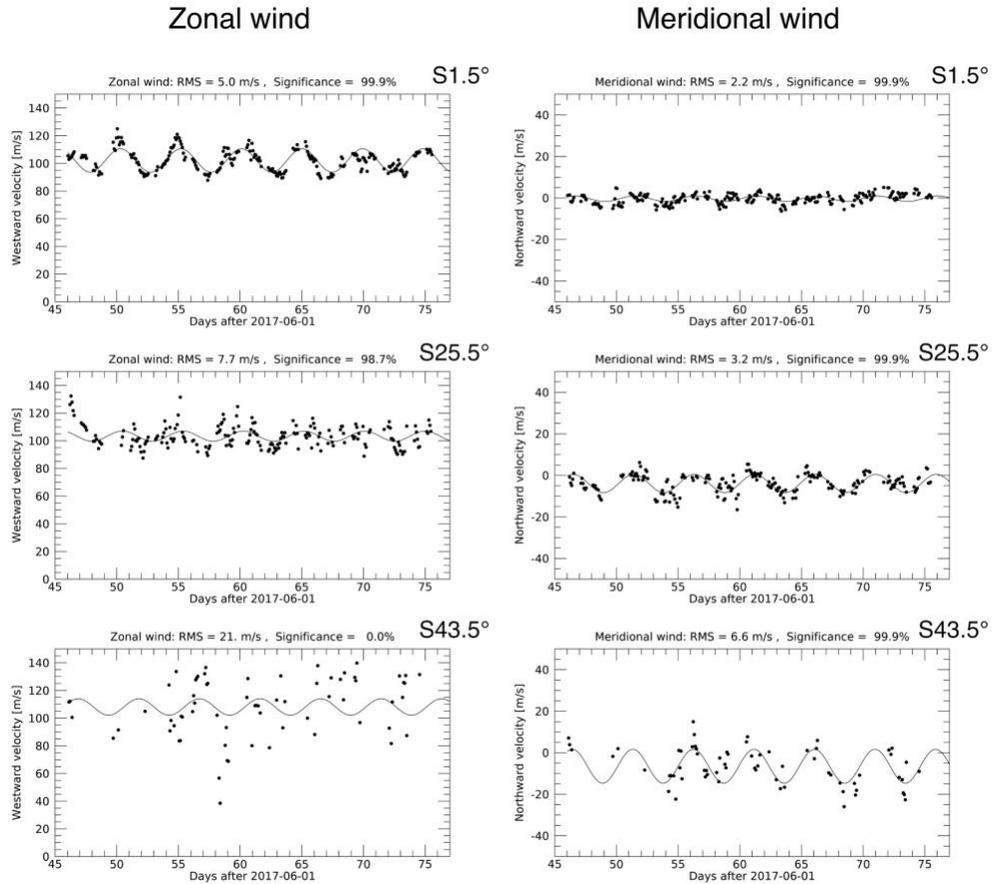

**Figure 19.** Time series of the zonal (left) and meridional (right) winds, as well as best-fit sinusoidal curves with a 5.1-day periodicity. Three latitude regions (1.5°S, 25.5°S, and, 43.5°S) are shown. The RMS and significance of the fitted curves are shown at the top of each panel.